\magnification1200


\vskip 2cm
\centerline
{\bf   Irreducible representations of E theory}
\vskip 1cm
\centerline{  Peter West}
\centerline{Department of Mathematics}
\centerline{King's College, London WC2R 2LS, UK}
\vskip 2cm
\leftline{\sl Abstract}
We construct the E theory analogue of the particles that transform under the Poincare group, that is, the irreducible representations of the semi-direct product of the Cartan involution subalgebra of $E_{11}$ with its vector representation. We show that one such irreducible representation has only the  degrees of freedom of eleven dimensional supergravity. This representation is most easily discussed in the light cone formalism and we show that the duality relations found in E theory take a particularly simple form in this formalism. We explain that the mysterious symmetries found recently in the light cone formulation of maximal supergravity theories are part of $E_{11}$. We also argue that our familiar spacetimes have to be extended by additional coordinates when considering extended objects such as branes.

\vskip2cm
\noindent

\vskip .5cm

\vfill
\eject

\medskip
{\bf 1. Introduction}
\medskip
Some time ago it was conjectured that the low energy effective action of strings and branes had an $E_{11}$ symmetry; the low energy effective action being   the non-linear realisation of  the semi-direct product of $E_{11}$ with the vector  ($l_1$) representation, denoted by $E_{11}\otimes_s l_1$ [1,2]. More recently it was shown that the essentially unique equations of motion 
which follow from this non-linear realisation contained the maximal supergravity theories in eleven and five dimensions [3,4]. There is good evidence that the non-linear realisation, for different decompositions of $E_{11}$,  contain all the maximal supergravity theories and also all their extensions to gauged supergravities, see reference  [5] for a review. 
\par
 The Poincare algebra in four dimensions is the semi-direct product of the Lorentz algebra, SO(1,D-1),   with the four translations $T^4$ and we may write it as $SO(1,D-1)\otimes_s T^4$. In 1939 Wigner derived all the irreducible representations of the Poincare algebra using the method of induced representations which is better known to physicists as the little group method [6]. These irreducible representations are just the particles that appear in relativistic quantum field theory. The irreducible representations  can be embedded into covariant representations of the Poincare group and the embedding conditions are the familiar on shell conditions for the linearised particle. The analogue of the Lorentz  algebra in the $E_{11}$ context is the Cartan involution invariant subalgebra of $E_{11}$,  denoted $I_c(E_{11})$. In the decomposition suitable to find the eleven dimensional theory  $I_c(E_{11})$ is, at lowest level,  just the Lorentz algebra. The analogue of the Poincare algebra is  $I_c(E_{11})\otimes_s l_1$ and at lowest level this is indeed just the Poincare group.  In this paper we will apply the method of the induced representation to find   the irreducible representations of  $I_c(E_{11})\otimes_s l_1$. There are very many such representations but we will discuss a few  of the interesting ones in some detail.  
 \par
 We now briefly recall some of the properties of $E_{11}$ that we will need in this paper and this will also establish the notation that we are using. A detailed account can be found in the book [7]. The positive  level generators of $E_{11}$ in its  decomposition to SL(11) are given by [1]
$$ K^{\underline a}{}_{\underline b}; \ R^{\underline a_1\underline a_2\underline a_3}; \ R^{\underline a_1\ldots \underline a_6}; \ R^{\underline a_1\ldots \underline a_8,\underline b} , \ldots  
\eqno{(1.1)} $$
and the  negative definite level generators of $E_{11}$  by:
Ê$$ R_{\underline a_1\underline a_2\underline a_3} ; \ R_{\underline a_1\ldots \underline a_6} ; \ R_{\underline a_1\ldots \underline a_8,\underline b} , \ldots 
\eqno{(1.2)} $$
The Cartan involution takes the negative roots to the positive roots and vice-versa. Its action on the  i$E_{11}$  generators are given by 
$$ I_c(K^{\underline a}{}_{\underline b}) = -K^{\underline b}{}_{\underline a}, \ \ \ I_c(R^{\underline a_1\underline a_2\underline a_3}) = -R_{\underline a_1\underline a_2\underline a_3}, $$
$$ I_c(R^{\underline a_1\ldots \underline a_6}) = R_{\underline a_1\ldots \underline a_6}, \ \ \ I_c(R^{\underline a_1\ldots \underline a_8,\underline b}) = -R_{\underline a_1\ldots \underline a_8,\underline b} \ldots  
\eqno{(1.3)} $$
As a result the Cartan involution invariant subalgebra is given by 
$$
I_c(E_{11}) = \{ J_{\underline{a}_1 \underline{a}_2}, S_{\underline{a}_1 \underline{a}_2 \underline{a}_3}, S_{\underline{a}_1 \ldots \underline{a}_6}, S_{\underline{a}_1 \ldots \underline{a}_8, \underline{b}}, \ldots \}
\eqno(1.4)$$
where 
$$
 J_{\underline a \underline b}=K^{\underline c}{}_{\underline b} \eta_{\underline a\underline c}- \eta_{\underline b\underline c}K^{\underline c}{}_{\underline a} ,\  S^{\underline a_1\underline a_2\underline a_3}=  R^{\underline a_1\underline a_2\underline a_3} - \eta ^{\underline a_1\underline b_1}\eta^{\underline a_2\underline b_2}\eta^{\underline a_3\underline b_3}R_{\underline b_1\underline b_2\underline b_3}, \ldots 
\eqno(1.5)$$
\par
An infinitesimal element of $I_c(E_{11})$ can be written in the form 
$$
h = I - (\Lambda^{\underline{a} \, \underline{b}} J_{\underline{a} \,  \underline{b}} + \Lambda^{\underline{a_1} \,  \underline{a_2} \, \underline{a_3}} S_{\underline{ a_1} \, \underline{a_2} \, \underline{a_3}} + \Lambda^{\underline{a_1} \ldots \underline{a_6}} S_{\underline{a_1} \ldots \underline{a_6}} + \Lambda^{\underline{a_1} \ldots \underline{a_8}, \underline{a}_9} S_{\underline{a_1} \ldots \underline{a_8},\underline{a}_9}  + \ldots ) 
\eqno(1.6)$$
\par
The elements of the vector representation are given by 
$$
l_1 = \{ P_{\underline{a}}, Z^{\underline{a}_1 \underline{a}_2}, Z^{\underline{a}_1 \ldots \underline{a}_5}, Z^{\underline{a}_1 \ldots \underline{a}_8}, Z^{\underline{a}_1 \ldots \underline{a}_7,\underline{b}}, \ldots \}, 
\eqno(1.7)$$
and their transformations under the group element of equation (1.6) are given in appendix A. 

\medskip
{\bf 2. Light cone analysis of field strengths}
\medskip
In this section we will examine how the duality relations found in E theory appear in the light cone formalism. This will be useful when we analyse,  in section 5.2,  the irreducible representation of $I_c(E_{11})\otimes_s l_1$ which is the analogue of usual the massless particle.  Given any vector $V^a , a=0,1,\ldots , D-1$,  we take its components in the  light cone to be 
$$
V^\pm= {1\over \sqrt 2} (V^{D-1}\pm V^0) ,\ \ V^i , \ i=1,\ldots ,D-2
\eqno(2.1)$$
We note that the non-zero components of the metric are $\eta_{+-}=1$ and $\eta_{ij} =\delta_{ij}$ and so $V_\pm= {1\over \sqrt 2} (V_{D-1}\pm V_0) $. 
\par
A light cone analysis of the irreducible representations of the Poincare group, that is, the on shell states have been analysed in reference [32] and [33] with an emphasis on the role played by the gauge invariant  field strengths. This included the introduction of gauge potentials in the  Lorentz "rest" frame of the particles. There is also a large literature on putting theories in the light cone formalism. These  usually decompose the gauge field into light cone components and then analyse the consequences of the equations of motion after making a gauge choice. In this paper we will analyse the field strengths  and gauge fields in the light cone components without making any choice of gauge or Lorentz frame. However, our analysis of the field strengths closely follows that in references [32] and [33]. 
\medskip 
{\bf 2.1 Form fields}
\medskip
We first consider the three from field whose field strength $F_{a_1\ldots a_4}\equiv  4\partial_{[a_1} A_{a_2a_3a_4]}$ which obeys the Bianchi identity  
$$
\partial_{[a_1} F_{ a_2\ldots a_5 ]}=0
\eqno(2.2)$$
as well as the  equation of motion 
$$
\partial^b F_{b a_1\ldots a_3}=0= \partial_{+}F_{- a_1\ldots a_3}+ \partial_{-}F_{ + a_1\ldots a_3}+ \partial^i F_{i a_1\ldots a_3}
\eqno(2.3)$$
\par
We will  take $\partial_{-}$ to be invertible. This  corresponds in momentum space to $p_{-}$ being non-zero. In this case   we can solve for  the $i_1i_2i_3$ and $- i_1i_2$ components of the equation of motion as follows 
$$
F_{+i_1i_2i_3}=-{1\over \partial_{-}}(\partial_+ F_{-i_1i_2i_3}+\partial^j F_{j i_1i_2i_3})
\eqno(2.4)$$
and 
$$
F_{+ - i_1i_2}={1\over \partial_{-}}\partial^j F_{- j i_1i_2}
\eqno(2.5)$$
The equation of motion with the indices $+i_1i_2$ is then automatic satisfied. Hence we have solved all the components of the field strength in terms of $F_{-i_1i_2i_3}$ and $F_{i_1i_2i_3i_4}$. 
\par
Taking the $-i_1i_2i_3i_4$ component of the Bianchi identity of equation (2.2)  we can solve for the $F_{i_1i_2i_3i_4}$ to find 
$$
F_{i_1i_2i_3i_4}={4\over \partial_{-}} \partial _{[ i_1 |}F_{- | i_2i_3i_4]}
\eqno(2.6)$$ 
Hence for all the components of the field strength $F_{a_1a_2a_3a_4}$ can be expressed in terms of $F_{-  i_1i_2i_3}$ which we can regard as our independent degrees of freedom. Taking the derivative $\partial^{a_1}$ of the Bianchi identity of equation (2.2) and using equation (2.3) we find that 
$\partial^2 F_{a_1a_2a_3a_4}=0$ and so 
$$
\partial^2 F_{-i_1i_2i_3}=0 
\eqno(2.7)$$ 
\par
We can also introduce the gauge fields $A_{a_1a_2a_3}$. We observe that  
$$
F_{-i_1i_2i_3}= \partial_{-} A_{i_1i_2i_3}- 3\partial_{[i_1 |}A_{- |i_2i_3 ]}= \partial _{-} \hat A_{i_1i_2i_3}
\eqno(2.8)$$
where $ \hat A_{i_1i_2i_3}=   A_{i_1i_2i_3}-3{1\over \partial_{-} } \partial_{[i_1 |} A_{- | i_2i_3]}$ is gauge invariant. Equation (2.7)  then implies that 
$\partial^2 \hat A_{i_1i_2i_3}=0$
The above analysis applies to any form field with very minor modifications. As advertised, by starting from  the field strength in the  light cone  rather than the gauge fields we have a procedure that  is completely gauge invariant and requires no gauge choice.
\medskip
{\bf 2.2 Gravity}
\medskip
Let us apply the same approach to linearised gravity. The Riemann tensor $R_{a_1a_2 , b_1b_2}= R_{ b_1b_2, a_1a_2 }$ obeys the Bianchi identities 
$$
\partial _{[ a_1 } R_{a_2a_3 ] , b_1b_2}=0,\ \ {\rm and }\  \ \partial _{[ b_1 |} R_{a_1a_2  , | b_2b_3 ]}=0
\eqno(2.9)$$
Tracing on $a_1$ and $b_1$ in  equation (2.9) and using the  equation of motion $R_{a , }{}^b\equiv R_{a c, }{}^{bc}=0$ implies that 
$$
 \partial^c R_{c a , b_1b_2}=0  ,\ \ {\rm and }\  \ \partial ^{ c } R_{a_1 a_2 ,  c b }=0
\eqno(2.10)$$
As a result acting with $\partial^{a_1}$ on equation (2.9) and using equation (2.10) we conclude that $\partial^2 R_{a_1a_2 , b_1b_2}=0$. 
Using the same steps as  we did  for the three form we can solve the equations (2.10). We find that 
$$
R_{+i , b_1 b_2 }= -{1\over \partial_{-}} (\partial_{+} R_{-i, b_1 b_2 } +\partial^j R_{ji, b_1 b_2 } ),\  \  R_{a_1a_2 , + i  }= -{1\over \partial_{-}} (\partial_+ R_{a_1a_2, - i  } + \partial^j  R_{a_1a_2 , j i  })
\eqno(2.11)$$
and  
$$
R_{+- , b_1 b_2 }= {1\over \partial_{-}} \partial^j R_{- j , b_1 b_2 }  ,\   \  R_{a_1a_2 , + -  }= {1\over \partial_{-}} \partial^j R_{a_1a_2, - j  } 
\eqno(2.12)$$
The Bianchi identity of equation (2.9) implies that 
$$
R_{ij, b_1b_2}= {2\over \partial_{-}}  \partial_{[i | } R_{- | j] , b_1b_2}  ,\ \ {\rm and }\  
R_{a_1 a_2, ij }= {2\over \partial_{-}}  \partial_{[i |} R_{a_1a_2, -| j]}
\eqno(2.13)$$
Using  equations (2.11), (2.12) and (2.13) we can solve for all the components of the Riemann curvature in terms of 
$R_{-i , -j}$ which we can  take to be the independent degrees of freedom. It obeys the equations 
$$
R_{--}= R_{-i , -i}=0 , \ \ {\rm and }\ \ \partial^2 R_{-i , -j}=0
\eqno(2.14)$$ 
We recognise that $R_{-i , -j}$ has the correct number of degrees of freedom 
to describe gravity. 
\par
We now recall the well known expressions for the Riemann tensor in terms of the linearised veilbein. 
The linearised spin connection is defined  as
$$
 \omega_{a}{}_{, b_1b_2}= - G_{b_1, b_2 a} + G_{b_2, b_1 a}+G_{a, b_1b_2}\equiv -\partial_{b_1}h_{(b_2 a)}+ \partial_{b_2}h_{(b_1 a)}
 + \partial_a h_{[b_1b_2]}
 \eqno(2.15)$$
and the Riemann curvature is given by 
$$
R_{a_1a_2 ,}{}^{ b_1b_2}= 2 \partial_{[a_1} \omega_{a_2]}{}_{, b_1b_2}= -4 \partial_{[a_1 } \partial^{[b_1} h^s_{a_2]} {}^{b_2]}
\eqno(2.16)$$
where $h^s_{ab}= h_{(ab)}$. 
Evaluating the component of the independent Riemann tensor in terms of the linearised vielbein we find that 
$$
R_{-i , -j}= -\partial_{-}^2 \hat h _{ij}
\eqno(2.17)$$
where $\hat h_{ij}= h_{ij} +\partial_i \chi_j +\partial_j\chi_i $ and $\chi_i= -{1\over \partial_{-}}h_{-i} + {1\over 2\partial_{-}^2}\partial_ih_{--}$.
It is easy to see that $\hat h_{ij}$ is  invariant under diffeomorphisms. 
 It follows from equation (2.14)that $\hat h_{ij}$ obeys the constraints $\hat h_i{}^i=0=\partial^2 \hat h_{ij}$. 
\par
Our independent degrees of freedom, $R_{-i , -j}$,  are given in terms of the spin connection  by 
$$
R_{-i , -j}= \partial_- \omega _{i, -j}-\partial_i \omega _{-, -j}
\eqno(2.18)$$
Using equation (2.15) we find that 
$$
\omega _{i, -j}= -\partial_{-} \hat h_{ij} + \partial_{i}\Lambda _{-j} , \ \ {\rm and }\ \ \omega _{-, -j}= \partial_{-} \Lambda_ {-j} 
\eqno(2.19)$$
where the local  Lorentz transformation $\Lambda_ {-i} $ is given by $\Lambda_ {-i} = -h_{i-}+{\partial_i\over \partial_{-}}h_{--}$. 
\par
We can apply similar arguments to those given above,  and those in the previous section,   to any field strength $F_{\ldots , a_1\ldots a_p, \ldots }$ that has blocks of antisymmetrised indices that 
obey the equations 
$$
\partial_{[a_1 |}F_{\ldots , | a_2\ldots a_{p+1}], \ldots } =0
\eqno(2.20)$$
and 
$$
\partial^b F_{\ldots , b a_1\ldots a_{p-1}, \ldots }=0
\eqno(2.21)$$
Using the above arguments we  conclude that can solve for all the components in terms of $F_{\ldots , -i_1\ldots i_{p-1}, \ldots }=\partial_{-}\ldots \partial_{-}  \hat A_{\dots , i_1\ldots i_{p-1}, \ldots }$
where $ \hat A_{\dots , i_1\ldots i_{p-1}, \ldots }$ is gauge invariant.


\medskip
{\bf 3 Duality relations in the light cone formalism}
\medskip
In this section we will examine the duality relations found in E theory in eleven dimensions. In order to make easy contact with the literature on E theory we will use the same notation as used there. For simplicity we will work at the linearised level. 
The three from and six form in eleven dimensional supergravity obey, at the linearised level,  the relation 
$$
G_{a_1a_2a_3a_4}={1\over 2.4!} \epsilon _{a_1a_2a_3a_4}{}^{b_1\ldots b_7} G_{b_1\ldots b_7}
\eqno (3.1)$$
where  $G_{a_1a_2a_3a_4}= {1\over 4} F_{a_1a_2a_3a_4}=\partial_{[a_1} A_{a_2a_3a_4]} $ and 
$G_{b_1\ldots b_7}= \partial_{ [b_1} A_{b_2\ldots b_7]}$. More precisely $G_{a_1a_2a_3a_4}= G_{[a_1, a_2a_3a_4] }$ where the later quantity is the one that occurs in the E theory papers. Using the analysis of section two we find that the independent components are $G_{-i_1i_2i_3}$ and $G_{-j_1\ldots j_6}$ and these obey the duality relation 
$$
G_{-i_1i_2i_3}={7\over 2.4!} \epsilon _{-i_1i_2i_3}{}^{-j_1\ldots j_6} G_{-j_1\ldots j_6}
=-{7\over 2.4!} \epsilon _{i_1i_2i_3}{}^{j_1\ldots j_6} G_{-j_1\ldots j_6}
\eqno(3.2)$$
as we take $ \epsilon ^{+-j_1\ldots j_9} =  \epsilon ^{j_1\ldots j_9} $. Using equation (2.8), and its generalisation, this last equation can be written as 
$$
\partial_{-} \hat A_{i_1 i_2i_3}=-{1\over 12} \epsilon _{i_1i_2i_3}{}^{j_1\ldots j_6} \partial_{-} \hat A_{j_1\ldots j_6}
\eqno(3.3)$$
Removing $\partial_{-}$ we find that 
$$
\hat A_{i_1 i_2i_3}=-{1\over 12} \epsilon _{i_1i_2i_3}{}^{j_1\ldots j_6}  \hat A_{j_1\ldots j_6}
\eqno(3.4)$$
Hence the dual gauge fields are themselves  related by the alternating symbol. 
\par
Let us now consider the duality relation between the graviton and the dual graviton [1] 
$$
\omega _{a}{}^{b_1b_2}\dot ={1\over 4}  \epsilon^{b_1b_2 c_1\ldots c_9} G_{c_1\ldots c_9, a}
\eqno(3.5)$$
where at the linearised level $G_{c_1\ldots c_9, a}= \partial_{[c_1} h_{c_2\dots c_9], a}$ and the dot above the equals sign indicates that 
the above relation is modulo Lorentz transformations as explained in references [8],[9], [4] and [18]. 
\par
We found in section that  two gravity is described the independent variables $\hat h_{i, j}$ which appears in  equation (2.19). Applying the analysis to the dual graviton we conclude that its independent component can be taken to be  $\hat h_{i_1\ldots i_8, j}$ which appears in $G_{-i_1\ldots i_8, j}= \partial_{-} \hat h_{i_1\ldots i_8, j}$. For these components the duality relation of equation (3.5) reads 
$$
\partial_{-} \hat  h^s _{i, j}\dot = -{1\over 4}  \epsilon_{j}{}^{ k_1\ldots k_8} \partial_{-}  \hat h_{k_1\ldots k_8, i}, \ \ {\rm or}\ \ 
\hat h^s _{ij}= -{1\over 4}  \epsilon_{j }{}^{k_1\ldots k_8} \hat h_{k_1\ldots k_8, i},
\eqno(3.6)$$
The Lorentz transformation terms that occurs in the spin connection of equation (2.19) can be discarded as the duality relation is modulo Lorentz transformations. We observe that the constraint $\hat h^s_i{}^i=0$ implies $\hat h_{[i_1\ldots i_8, j]}=0$ and vice versa. We recall that in  E theory the dual graviton obeys the irreducibility condition $\hat h_{[a_1\ldots a_8, b]}=0$ and so the equation at the end of the last sentence is automatically true. 
\par
At level four in the non-linear realisation we find the field $A_{a_1\ldots a_9, b_1b_2b_3}$ which obeys a  duality relation with the three form field that is given by [18,11]
$$
G_{a_1\ldots a_{10}, b_1\ldots b_4}= {36\over 5. 11!}\epsilon_{a_1\ldots a_{10}}{}^{c}\partial_c G_{b_1\ldots b_4}
\eqno(3.7)$$
where $G_{a_1\ldots a_{10},}{}^{ b_1\ldots b_4}=\partial_{[ a_1}\partial^{[b_1}A_{a_2\ldots a_{10}],}{}^{ b_2\ldots b_4]}$ . 
From the duality relation we find that equations (2.20) and (2.21) hold for both blocks of indices on $G_{a_1\ldots a_{10},}{}_{ b_1\ldots b_4}$. Hence  
applying the  analysis of section two we conclude that the independent components of the field strength are  $G_{-i_1\ldots i_9, - j_1j_2j_3}$ and these  obeys the duality relation 
$$
G_{-i_1\ldots i_9, - j_1j_2j_3}= -{36\over 5. 11!}\epsilon _{i_1\ldots i_9} \partial_{-}G_{-j_1j_2j_3} 
\eqno(3.8)$$
which in turn implies that 
$$
\hat A_{i_1\ldots i_9,  j_1j_2j_3}=- {36\over 5. 11!} \epsilon _{i_1\ldots i_9} \hat A_{j_1j_2j_3} 
\eqno(3.9)$$
At higher levels in E theory we find fields with three indices and any number of blocks containing nine antisymmetrised indices and similarly 
for fields that have six indices. These obey duality relations, similar to the  one  we have given in equation (3.7). These relations were discussed in 
reference [11]. It is straight forward to analyse these in the light cone formalism given in this paper and we find that all the fields are related to the fields $\hat A_{j_1j_2j_3}$ by multiples of  $\epsilon _{i_1\ldots i_9}$. 
\par
We close this section by discussing the similar higher order dualities in the gravity sector. Here we find the graviton $h_{ab}$. the dual graviton field $h_{a_1\ldots a_8, b}$, whose duality relation we encountered above, and, at higher levels, the fields   $h_{a_1\ldots a_9, b_1\ldots b_8,c}$ and  
$h_{a_1\ldots a_9, b_1\ldots b_9, c_1\ldots c_8,d}$ etc. The  $h_{a_1\ldots a_8, b}$, and $h_{a_1\ldots a_9, b_1\ldots b_8,c}$ should obey the duality relation [10]
$$
G_{a_1\ldots a_{10}, b_1\ldots b_9,c_1c_2 }=e_1 \epsilon _{a_1\ldots a_{10} e}\partial^e G_{ b_1\ldots b_9, c_1c_2}
\eqno(3.10)$$
where $e_1$ is  a suitable chosen constant, whose value is fixed by higher order calculations in E theory, and 
$$ 
G_{ b_1\ldots b_9, }{}^{c_1c_2}=  \partial_{ [ b_1]}\partial^{[c_1}h_{b_2\ldots b_9],}{}^{ c_2]}  \  {\rm and  }\  
 G_{a_1\ldots a_{10}, b_1\ldots b_9,}{}^{c_1c_2 }= \partial_{[a_1}\partial _{[b_1 |}\partial^{[c_1}h_{a_2\ldots a_{10}],| b_2\ldots b_9],}{}^{c_2 ]}
\eqno(3.11)$$
One finds that equations (2.20) and (2.21) hold and so applying the above analysis we find that the independent components are given by 
$G_{-i_1\ldots i_{9}, -j_1\ldots j_8, - k}=\partial_{-}^3 \hat h_{i_1\ldots i_{9}, j_1\ldots j_8,  k}$ and  
$G_{-i_1\ldots i_8,-j}=\partial_{-}^2 \hat h_{i_1\ldots i_8,j} $ which obey the duality relation 
$$
G_{-i_1\ldots i_{9}, -j_1\ldots j_8, - k}=-e_1\epsilon_{i_1\ldots i_{9}} \partial_{-} G_{ -j_1\ldots j_8, - k}
\eqno(3.12)$$
and as a result 
$$
\hat h_{i_1\ldots i_{9}, j_1\ldots j_8, k}=-e_1\epsilon_{i_1\ldots i_{9}} \hat h_{ j_1\ldots j_8,  k}
\eqno(3.13)$$
\par
One can find the same result by first integrating  (3.10) to find the duality equation [10]
$$
\partial_{a_1} \hat h_{a_2\ldots a_{10}, b_1\ldots b_8,c}\dot =e_1 \epsilon _{a_1\ldots a_{10} e}\partial^e \hat h_{ b_1\ldots b_8, c}
\eqno(3.14)$$
which holds modulo certain gauge transformations. We can find the above results by starting from this relation and then apply similar arguments. The duality relations for the higher level fields in the gravity sector follow a similar pattern. 
\par
Thus we find that in light cone formalism  the fields that appear in the different dualities are essentially equal to each other once one uses the epsilon symbol. This is perhaps not so surprising as seen  from the view point of the induced representations discussed in the next section. Starting from a given irreducible representation of the Poincare group we can embed it into different covariant representations of the Lorentz group and as a result we required different projection conditions to the ensure that only the chosen irreducible representation is present. However, this latter irreducilbe representation is always the same. These lead to different on shell conditions on different fields and  so different equations of motion. The different choices of embedding fields are related by duality symmetries.  
\medskip
{\bf 4. The theory of irreducible representations of $I_c(E_{11})\otimes_s l_1$}
\medskip
The irreducible representations of the Poincare group were found in 1939 by Wigner [6]. Very briefly it went as follows; the group of Poincare  transformations that left invariant a given momentum was found. Then an irreducible  representation of this group, called the little group,  was chosen and  the full irreducible representation was obtained by carrying out a Lorentz boost. In this way all irreducible representations of the Poincare group were  found and so a description all possible particles. This method was formalised and it became known as the method of  induced representations. As we mentioned previously the Poincare algebra is the semi-direct product of the Lorentz algebra  with 
the translations and so it has the same structure as the algebra $I_c(E_{11})\otimes_s l_1$. Indeed the level zero part of this later algebra is the Poincare algebra in eleven dimensions. 
\par
We  will now explain how to construct the irreducible representation of $I_c(E_{11})\otimes_s l_1$ following  the same steps as was used for the Poincare algebra. The brane charges, denoted by $l_A$ are contained in the vector ($l_1$) representation of $E_{11}$ and we first select certain of these to be   non-zero. Denoting  these by $ l_a^{(0)}$,  we compute the subalgebra of $I_c(E_{11})$, denoted $\cal {H}$,  which leave these non-zero brane charges  invariant. Clearly for different choices of non-zero brane charges we find different subalgebras  $\cal {H}$. The variation of the brane charges under $I_c(E_{11})$ transformations  are given in appendix A. 
\par
We now choose an irreducible  representation of ${\cal {H}}\otimes_s l_1$ acting on the fields $\psi_I (0)$ as follows 
$$
U(L_A )\psi_I (0) =  l_A^{(0)}\psi(0)_I, \ \ {\rm and }\ \ U(h) \psi(0)_I= \hat D(h^{-1})_I{}^J\psi_J(0),\ \ h\in \cal {H}
\eqno(4.1)$$
where $L_A$ are the generators in the vector representation,  $U(L_A )$ and $U(h) $ denote the action of the corresponding generators on $\psi_I (0)$ and $\hat D$ is the chosen irreducible representation of ${\cal { H}}$. 
\par
Any group element $g$ of  $I_c(E_{11})$ can be written in the form $g=e^{\varphi\cdot S}h$ for some $h\in {\cal H}$ and $S$ denote the generators of  $I_c(E_{11})$ not in ${\cal H}$. We define the elements of the full irreducible representation by carrying out a "boost" as follows 
$$
\psi _I(\varphi)\equiv U(e^{\varphi\cdot S})\psi_I(0)
\eqno(4.2)$$
Acting with a generator $L_A$ in the vector representation we find that 
$$
U(L_A) \psi_I (\varphi) = U(e^{\varphi\cdot S})U(e^{-\varphi\cdot S})U(L_A )U(e^{\varphi\cdot S})\psi_I(0)
$$
$$
=U(e^{\varphi\cdot S})U(e^{-\varphi\cdot S } L_A e^{\varphi\cdot S})\psi_I(0)
=U(e^{\varphi\cdot S})U(D(e^{\varphi\cdot S})_A{}^B L_B )\psi_I(0)
$$
$$ 
=D(e^{\varphi\cdot S})_A{}^B l_B(0) \psi (\varphi)
= l_A (\varphi) \psi (\varphi)
\eqno(4.3)$$
where $l_A (\varphi) = D(e^{\varphi\cdot S})_A{}^B l_B(0)$ and $D$ is the matrix of the vector representation of  $I_c(E_{11})$. 
\par
We must also  show how any group element $g_0\in I_c(E_{11})$ acts on the representation;
$$ 
U(g_0) \psi_I (\varphi)= U(g_0) U(e^{\varphi\cdot S})\psi_I(0)
= U(g_0 e^{\varphi\cdot S})\psi_I(0)
$$
$$
=U(e^{\varphi^\prime \cdot S}h_c )\psi_I(0)
= U(e^{\varphi^\prime \cdot S})U(h_c )\psi_I(0)
= \hat D(h_c^{-1})_I{}^J \psi_J (\varphi^\prime)
\eqno(4.4)$$
where we have used the equation 
$$
g_0 e^{\varphi\cdot S}= e^{\varphi^\prime \cdot S} h_c
\eqno(4.5)$$
\par
The functions we have constructed so far are in the $E_{11}$ analogue of momentum" space but we can carry out a generalised Fourier  transformation to find functions that  depend on the (generalised) spacetime, that is, on the coordinates $x^A$ which are in a one to one correspondence with the brane charges and so the vector representation.  To this end we define the 
functions 
$$
\tilde \psi_I (x^A) \equiv \int D\varphi \ e^{ l_A(\varphi) x^A }\psi_I(\varphi)=  \int D\varphi \ U(e^{ L_Ax^A }) \psi_I(\varphi)
= \int D\varphi e^{x^A D_A{}^B (e^{\varphi\cdot S}) l_B (0)}\psi_I(\varphi)
\eqno(4.6)$$
We note that the dependence of spacetime occurs in the combination $x^A D_A{}^B (e^{\varphi\cdot S}) l_B (0)=x^A  l_A (\varphi)$ and it is completely determined by the group theory. 

Under a transformation $g_0\in I_c(E_{11})$ this function transforms as 
$$
U(g_0) \tilde \psi_I (x^A) =  \int D\varphi \ U(g_0 e^{x^AL_A }e^{\varphi\cdot S} )\psi_I(0)
= \int D\varphi \ U(e^{ x^{A\prime}L_A }e^{\varphi^\prime\cdot S}h_c )\psi_I(0)
$$
$$
= \int D\varphi \ e^{x^{A\prime } l_B (\varphi^\prime )} \hat D(h^{-1}_c)_I{}^J\psi_J(\varphi^\prime)
\eqno(4.7)$$
where $g_0 e^{ x^AL_A }e^{\varphi\cdot S} =  e^{{x^A}^\prime L_A }e^{\varphi^\prime \cdot S} h_c$. The action of $L_A$ is just differentiation  by ${\partial\over  \partial x^A}$.
The integration over $\varphi$ can be converted into an integration over the "momenta" (brane charges) with suitable constraints. We observe that the relations between the group elements are very similar to those that occur in  the  the non-linear realisation of $E_{11}\otimes_s l_1$ used in the construction of brane dynamics [12],[13] and [14].
\par
The irreducible representation of  a group can be labelled by the values of its  Casimirs. For the Poincare group in four dimensions we have the Casimirs $P_a P_b \eta^{ab}$ and the Pauli Lubanski vector. Strictly speaking to systematically  find all the irreducible representation we should for each of the  possible  values of the Casimirs   first  choose some  momenta that result in  these value. After this step we should then   proceed as advocated at the beginning of this section, namely choose an irreducible representation of ${\cal H}$ acting on functions evaluated for these chosen momenta. For the case of $I_c(E_{11})\otimes_s l_1 $ we do not have a  full set of Casimirs although we  know one  of them [15]
$$
L^2\equiv  L_A L_B K^{AB}= P_a P^a  + 2Z^{a_1a_2}Z_{a_1a_2}+5! Z^{a_1\ldots a_5}Z_{a_1\ldots a_5}
$$
$$
+7! Z^{a_1\ldots a_8}Z_{a_1\ldots a_8}+ 9.7! Z^{a_1\ldots a_7, b }Z_{a_1\ldots a_7, b }+\ldots 
\eqno(4.8)$$
where $K^{AB}$ is the $I_c(E_{11})$ invariant metric. Clearly this is the  $I_c(E_{11})$ generalisation of $P_a P_b \eta^{ab}$. This was observed to vanish for all known half BPS invariant solutions. 
\par
As the Casimir  in equation (4.8) is  $I_c(E_{11})$ invariant,  its value is the same when applied to any state in the representation. As such its value on the irreducible representations discussed above  are just given by those on $\psi_I (0)$. Implementing this condition on  the functions of equation (4.6) leads to differential conditions which for the Casimir of equation (4.8)  takes the form 
$$
({\partial \over \partial x^a} {\partial \over \partial x^b}\eta^{ab} +2{\partial \over \partial x_{ a_1a_2} }{\partial \over \partial x_{ b_1b_2} }
\eta^{a_1b_1} \eta^{a_2b_2}+\ldots ) \tilde \psi_I (x^A)= L^2(0) \tilde \psi_I (x^A)
\eqno(4.9)$$
where $L^2(0)$ is the value on the Casimir on $\psi(0)$. We can think of this as a mass shell condition. 
\par
An interesting  $I_c(E_{11})$ multiplet was discuss in reference [16]. There one took the tensor product of two  vector representations, that is,  $l_1\otimes l_1$ and restricted it to belong to the fundamental representation associated with node ten, denoted $l_{10}$. At low levels this has the components 
$$
\Delta _{10}^{(2)} \equiv l_1\otimes l_1 |_{l_{10}}= (P_b Z^{ba} , \ Z^{[a_1a_2} Z^{a_3a_4]}+ P_b Z^{b a_1\ldots a_4}, \ 
   P_bZ^{ba_1\ldots a_7}   -3 Z^{[a_1a_2 }Z^{a_3\ldots a_7]} ,\ 
   $$
   $$
    P_c Z^{c a_1 \ldots a_6, b}+ {6.5.3 \over 7}( Z^{b [a_1} Z^{a_2\ldots a_6 ]}-   Z^{ [a_1 a_2} Z^{a_3\ldots a_6] b}) ,\ 
\ldots 
\eqno(4.10)$$
where $\ldots$ indicate the infinite number of higher level components. Unlike restricting to some other fundamental representations,  setting this multiplet to zero does not result in all the brane charges being zero.  This multiplet was also observed to vanish for the known half BPS solutions [15]. In fact the objects in equations (4.8) and (4.10) are $E_{11}$ generalisations of quantities considered in reference [16] in the context of the reduction of maximal supergravity to three dimensions which has  $E_{8}$ symmetry. 
\par
If  this $I_c(E_{11})$ multiplet vanishes on the functions $\psi_I(0)$ it will vanish on the entire irreducible representation. In this case it will happen the functions defined on the generalised spacetime will obey the conditions 
$$
{\partial \over \partial x^b} {\partial \over \partial x_{ ba}} \tilde \psi_I (x^A)= 0, \ \ldots 
\eqno(4.11)$$
We can think of this as an on-shell condition but only for a choice of brane charges that is "half BPS". We will see that there are interesting irreducible representations for which this is not the case. 
\par
Another  interesting  $I_c(E_{11})$ multiplet can be constructed out of the the tensor product of two $ \Delta^{(2)} _{10}$'s  restricted to the fundamental representation associated with node six, that is, $\Delta ^{(4)}_6\equiv \Delta^{(2)} _{10}\otimes \Delta ^{(2)}_{10} |_{l_{6}}$  [15]. This does not vanish for half BPS solutions but it does vanish for one quarter BPS solutions. We note that this  last multiplet is quartic in the vector representation. One can proceed as before to use this to  impose  conditions on the fields defined on the generalised spacetime when they concern a "quarter BPS' brane charges. One could continue in this manner to construct a higher order invariant out of $\Delta ^{(4)}_6$. 
\par
Before proceeding further, the reader may appreciate a very brief reminder of how the next steps go for the case of the Poincare group in four dimensions.  A review is given in reference [17]. The simplest case is for a massive particle,   $p^2=-m^2$ and we can choose $p_0=m$,  all other momenta being zero. This choice is preserved by  the subalgebra ${\cal H}= \{J_{ij}\}$ with $i,j,\ldots =1,2,3$ and so the remaining  generators in the Lorentz algebra are $J_{0i}$. We must now choose a representation of ${\cal H}$ which,  for simplicity,  we choose to be the vector $\psi_i(0)$. The full representation is found by boosting, using equation (4.2) to define  $\psi_i(\varphi) = exp (\varphi^{0i} J_{0i}) \psi_i(0)$. Using equation (4.6) the fields in $x$ spacetime are given by 
$$
\int d\varphi e^{x^aP_a} \psi_i (\varphi) = \int d\varphi e^{x^a\Lambda_a{}^0(\varphi) m } \psi_i (\varphi)
\eqno(4.12)$$
where $\Lambda_a{}^0(\varphi)= D_a{}^0 (e^{\varphi^{0k} J_{0k}})$ is the Lorentz transformation induced by the boost and $d\varphi = d\varphi^{0i}$. 
We can change the integration from over $\varphi^{0i}$  to be over $p_i$ and finally to be over the more familiar $d^4 p \delta(p^2+m^2)$ with suitable factors. 
\par
To find a covariant formulation we enlarge the representation to the four vector $ A_a(0) $ such that 
$ A_i(0) =\psi_i(0)$  and $A_0(0)=0$. Boosting these conditions we find that they are equivalent to  $p^a A_{a}=0$. We also have  the Casimir condition $p^2+m^2=0$. Thus the irreducible representation in  x space is described by the field $A_a(x)$ subject to the conditions  
$$(\partial^2 +m^2)A_a(x)=0 ,\ \ {\rm and }\ \  \partial^a A_a(x)=0, 
\eqno(4.13)$$ 
which we recognise as the correct conditions for a massive spin one. 
\par
We now recall the treatment of a massless vector as an irreducible representation of the Poincare group. In this case  $p^2=0$ and we can choose $p_0=-m$  and $p_3=m$ with $p_1=0=p_2$, or equivalently,  $p_{-}=\sqrt 2 m$, with  all other components of the momenta being zero. The algebra that preserves this choice is 
${\cal H}=(J_{+i}, \  J_{ij})$ where now $i,j \ldots =1,2$. However, $[ J_{+i}, J_{+j}]=0$ and  in order to find a finite dimensional unitary representation we must take $J_{+i}$ to vanish on our representation. As  a result we are required to  take  an irreducible representation of just the $\{J_{ij}\}$. We will choose this to be $\psi_i(0)$. The generators outside ${\cal H}$  are  $J_{-i} , \ J_{+-}$ and, following equation (4.2)  we find the full representation by boosting with these and so it is parameterised by 
$\varphi^{-i} , \ \varphi^{+-}$. The corresponding x space fields that describe the irreducible representation can be found  by applying equation (4.6) to the case of the Poincare algebra. In this we may swop the integration over the three $\varphi$'s for one over $d^4 p\delta (p^2=0)$ with suitable factors. 
 The x space dependence in  the analogue of formula (4.6)  is $x^a D_a{}^- (exp (\varphi^{-i} J_{-i}+ J_{+-} \varphi^{+-}) ) \sqrt 2m$  where 
 $D_a{}^- (exp (\varphi^{-i} J_{-i}+ J_{+-} \varphi^{+-}) $ is the Lorentz transformation induced by the boost.  
 \par
 What is much more subtle in the massless case  is the embedding of the irreducible representation into a representation of the Lorentz group. We take this to be the vector $A_a(0)$, such that $A_i(0)=\psi_i(0)$ for $i=1,2$, $A_{+}(0)=0$ but $A_{-}(0) $ subject to the equivalence relation $A_{-}(0)\sim A_{-}(0) + \Lambda(0)$. These conditions are equivalent to $p^a A_a(0)=0$ and $A_a(0)\sim A_a(0)+p_a\Lambda (0)$. After  the boost field in x space these  conditions can be written as $\partial^a A_a(x)=0$ and 
$A_a  (x) \sim A_a (x)  +\partial_a \Lambda (x) $ and we also have the Casimir condition $\partial^2 A_a (x)=0$. Thus the irreducible representation is described by the fields $A_a  (x)$ subject to the on-shell conditions 
$$
 \partial^2 A_a (x)=0 ,  \ \ \partial^a A_a  (x)=0 ,\ \ {\rm and \ the \  equivalence \ relation }\ \  
 A_a (x)\sim A_a  (x) +\partial _a \Lambda  (x)
 \eqno(4.14)$$
The conditions of equation (4.14) are not the same as the equations of motion which is given by    
$\partial^b (\partial_b A_a - \partial_a A_b)=0$. A formula which we recognise from our undergraduate days.  To get to the on-shell conditions we can choose the Lorentz gauge $\partial^a A_a  (x)=0 $ and recognise that we have a residual gauge symmetry. The field equations for any irreducible representation of the Poincare group were found in reference [33]. 
\par
 It is these steps that we must repeat for the irreducible representations of  the algebra $I_c(E_{11})\otimes_s l_1$. 
 Hence having  chosen an irreducible representation $ \psi_I (\varphi) $ of ${\cal H}$ we must  embed it into  a representation of  $I_c(E_{11})$ and discover the projection conditions of this embedding. In addition in the massless case we must also introduce equivalence relations.  We will do this explicitly for one of the irreducible representations in section (5.2). We observe that the above procedure for finding the irreducible representations gives an explicit expression for the dependence of the fields on the generalised spacetime. However, this is the dependence corresponding to  the on shell fields is  not the same as for the fields that obey the equations of motion, this is especially the case when  gauge symmetries are  present. 
\medskip
{\bf 5. Examples  of irreducible representations of $I_c(E_{11})\otimes_s  l_1$}
\medskip
In this section we will construct some of the irreducible representations of $I_c(E_{11})\otimes_s l_1$. As we do not have a systematic knowledge of the Casimirs we will simply choose some  interesting values,  $l_A(0)$,  for the brane  charges in the vector representation, find the subalgebra that preserves  this choice and then, in two  cases, choose  an  irreducible representation acting on the function evaluated for these brane charges. 
\medskip
{\bf 5.1 The massive case}
\medskip
We take all the charges in the vector representation to be zero except for $P_0=m$. It is the analogue of the usual massive particle. The Casimir $L^2$ of equation (4.8) has the value $-m^2$ and is non-zero and so   equation (4.9) holds with this value. 
Varying our chosen brane charges under the  $I_c(E_{11})$ variations given in appendix A we find that 
$$
\delta P_a = 2 m \Lambda^0{}_a , \ 
\delta Z^{b_1 b_2} = - 6m \Lambda^{0 b_1 b_2} ,\ 
\delta Z^{b_1 \ldots b_5} = 360 m \Lambda^{0 b_1 \ldots b_5},
$$
$$
\delta Z^{b_1 \ldots b_8} = {9 \cdot 6720 \over 8} m \Lambda^{b_1 \ldots b_8,0}, \ 
\delta Z^{b_1 \ldots b_7,d} = - 7560 m \Lambda^{b_1 \ldots b_7 d,0} - 60480 m \Lambda^{0 b_1 \ldots b_7,d}, \ \ldots 
\eqno(5.1.1)$$
The algebra that preserves our choice of brane charges,  $l_A(0)$ requires us to set these variations to zero  and so we find the parameter 
 $\Lambda^{\underline{\alpha}}$ in the group element of equation (1.6) must obey the   constraints 
$$
\Lambda^0{}^b = 0=\Lambda^{0 b_1 b_2} =\Lambda^{0 b_1 \ldots b_5}=\Lambda^{0 b_1 \ldots b_7,d}=\Lambda^{b_1 \ldots b_8,0}=  \ \ldots 
\eqno(5.1.2)$$
We notice that every parameter that has a $0$ index is zero. 
Hence we  conclude that the preserved subalgebra 
${\cal H}$ contains generators that contain no $0$ index and as a result the preserved subgroup  is 
$$
{\cal H}= I_c(E_{10})
\eqno(5.1.3)$$
\par
The next step in constructing the irreducible representation is to choose a representation of $I_c(E_{10})$. One such representation is provided by the generators of $E_{10}$ that are odd under the Cartan involution.  This follows from the fact that the commutators of  the even generators 
of $E_{10}$, that is,  $I_c(E_{10})$ with these odd generators result in the  odd generators.  The fields in this representation are 
$$
h_{ab}= h_{(ab)},\  A_{a_1a_2a_3}, \ A_{a_1\dots a_6}, \ A_{a_1\ldots a_8, b }, \ldots 
\eqno(5.1.4)$$
where the indices take the range $\ a,b,\ldots = 1,2, \ldots , 10$ and the fields are evaluated at $\varphi=0$. 
\par
 We can then boost, using equation (4.2),  to find the full representation and then make the transition to x space using equation (4.6). 
The  fields in the resulting irreducible representation will obey equation  (4.9) with $L^2(0)=-m^2$. Also the conditions of equation (4.11) will hold. 
\par
To obtain an $I_c(E_{10})$  covariant approach we need to embed the representation of equation (5.1.4) into a representation of $I_c(E_{11})$.  The most natural way to do this is to consider the Cartan involution odd generators of $E_{11}$ which do form a representation of $I_c(E_{11})$. Denoting the elements of this representation by fields we find that they   are given by equation (5.1.4) but now with the index range extended to be  $\ a,b,\ldots = 0, 1,2, \ldots , 10$. Clearly, for the fields evaluated for the chosen brane charges the projection condition is just that any field with a zero index vanishes. One way to covariantly implement this condition is to adopt the gauge fixed constraint of reference [21] and discussed in more detail in the next section. It is the analogue of $\partial^a A_a=0$ for the Poincare algebra case. 
\par
It would be interesting to explore in detail what are the physical degrees of freedom in the above irreducible representation and what are the equations of motion. One of the first points to resolve is if there are any invariant  relations between different fields. This is the case of the irreducible representation of the next section whose fields  satisfy duality relations so reducing the number of degrees of freedom. However, we recall that a duality equation relates a Bianchi identity to an equation of motion, once one takes a derivative. As a result duality relations are   closely tied to gauge symmetries which are usually absent for massive fields. 
 \medskip
 {\bf 5.2 The massless  case}
\medskip
We will now  take all the brane charges in the vector representation to be zero except for $p_0=-m$ and $ p_{10}=m $. In the light cone notation this  is  the same as $p_{-}=\sqrt 2 m$ with all other brane charges  vanishing. This choice for the Poincare group would correspond to a massless particle. The Casimir $L^2$ of equation (4.8) vanishes and the conditions of equation (4.11) hold and so 
from this point of view one can think of this as "half BPS".  Varying the chosen non-zero charges  under an  $I_c(E_{11})$ transformations of appendix A we find that the parameters are restricted by the conditions 
$$
\Lambda_{+a}=0= \Lambda _{+-}= \Lambda_{+a_1a_2 }= \Lambda_{+a_1\ldots a_5 }= \Lambda_{+a_1\ldots a_b }= \Lambda_{+a_1\ldots a_7,b }, \ldots 
\eqno(5.2.1)$$
where  the indices $a, a_1,\ldots $ can take any value. In other words any parameter with a lower $+$ component vanishes. The subalgebra ${\cal H}$ that leaves the chosen charges invariant  has the generators  
$$
{\cal H}=\{J_{+i}, \ J_{ij}, S_{+ij},  S_{i_1i_2i_3} ,\ S_{+i_1\ldots i_5} \ \ldots  \} , \ \ i,j,\ldots = 2,\ldots 10
\eqno(5.2.2)$$
which consists of all generators that have no $-$ index.  
\par 
It is straight forward to compute the algebra of ${\cal H}$ at low levels and we find that 
$$
[J_{+i}, J_{+j}]=0, \  [S_{+i_1i_2} , S_{+j_1j_2}]=0 , \  [J_{+i} , S_{+j_1j_2}]=0,\ \ldots 
\eqno(5.2.3)$$
while the generators with no $+$ index obey the  $I_c(E_{11})$ commutators with the indices restricted over the index range of $i,j ,\ldots = 2,\ldots 10$, that is,  the algebra $I_c(E_9)$. 
 \par
In the case of the Poincare algebra, discussed above,  we demanded that we take  a finite dimensional unitary representation and this requires us to take $J_{+i}$ to be trivially realised leaving only the generators  $ J_{ij}$.  We see from equation (5.2.3) that all the low level generators with a $+$ index have  commutators that vanish. We will assume that this holds at higher levels  and that we should take all these generators to be trivially realised. As a result the only surviving generators in ${\cal H}$  have the indices $i,j,\ldots =2,\ldots 10$ and so we are left with the Cartan involution invariant subalgebra of $E_9$, denoted $I_c(E_9)$. We recall that $E_9$ is the  affine extension of $E_{8}$ and that  $I_c(E_{8})=SO(16)$.
The upshot is that  we must look for  an irreducible representation of $I_c(E_9)$. 
\par
Although we could consider the algebra $E_9$ from the viewpoint of it being an affine algebra it will be advantageous to construct it in the way that it aroseabove  from 
$E_{11}$. This is achieved by taking generators in  $E_{11}$ that have   indices which are restricted to take only the values  $i,j,\ldots = 2, \ldots , 10$.  As a result  the generators in $E_{11}$ that have  blocks of completely antisymmetrised ten and eleven indices are excluded and we are left with the remaining generators.  Fortunately all these  generators were found in reference [29] and they have a very simple form: 
$$
\ldots , \ K^i{}_j, \ R^{i_1i_2i_3} ,\ R^{i_1\ldots i_6} ,\  R^{i_1\ldots i_8,j } ,\  \ R^{j_1\ldots j_9, i_1i_2i_3} ,\ R^{j_1\ldots j_9, i_1\ldots i_6} ,\
$$
$$
R^{j_1\ldots j_9, i_1\ldots i_8,j } ,\ R^{j_1\ldots j_9,k_1\ldots k_9, i_1i_2i_3} ,\ R^{j_1\ldots j_9,k_1\ldots k_9,  i_1\ldots i_6} ,\
R^{j_1\ldots j_9, k_1\ldots k_9, i_1\ldots i_8,j } ,\ \ldots 
\eqno(5.2.4)$$
where the $\ldots$ at the beginning of the list denotes the presence of the   negative level generators which mirror the positive level generators and the  $\ldots$ at the end  of the list denotes the presence of the higher level generators. Of course we can replace any of the blocks of nine indices by $\epsilon_{i_1\ldots i_9}$ and then we recognise the fact that the $E_9$ algebra is the affine  extension of $E_8$.  
\par
Given   the $E_{9}$ generators in   equation (5.2.4),  it is straightforward to construct its  Cartan involution invariant subalgebra $I_c(E_9)$ which has  the generators 
$$
J_{ij} , \ S_{i_1i_2i_3} ,\ S_{i_1\ldots i_6} ,\  S_{i_1\ldots i_8,j } ,\  \ S_{j_1\ldots j_9, i_1i_2i_3} ,\ S_{j_1\ldots j_9, i_1\ldots i_6} ,\
$$
$$
S_{j_1\ldots j_9, i_1\ldots i_8,j } ,\ S_{j_1\ldots j_9,k_1\ldots k_9, i_1i_2i_3} ,\ S_{j_1\ldots j_9,k_1\ldots k_9,  i_1\ldots i_6} ,\
S_{j_1\ldots j_9, k_1\ldots k_9, i_1\ldots i_8,j } ,\ \ldots 
\eqno(5.2.5)$$
where 
$$
J_{ij} = K^k{}_j\eta_{ki} - K^k{}_i\eta_{kj}, S_{i_1i_2i_3}= R^{j_1j_2j_3}\eta_{i_1j_1} \eta_{i_2 j_2} \eta_{i_3j_3} - R_{i_1i_2i_3}, \ \ldots 
\eqno(5.2.6)$$
\par
We now must choose an irreducible representation of  ${\cal H}$,  actually $I_c(E_9)$. We note that the Cartan involution odd generators of $E_{9}$ form a representation of $I_c(E_9)$; they  have the form 
$$
T_{ij}= T_{(ij)}, \ T_{i_1i_2i_3} ,\ T_{i_1\ldots i_6} ,\  T_{i_1\ldots i_8,j } ,\  \ T_{j_1\ldots j_9, i_1i_2i_3} ,\ T_{j_1\ldots j_9, i_1\ldots i_6} ,\
$$
$$
T_{j_1\ldots j_9, i_1\ldots i_8,j } ,\ T_{j_1\ldots j_9,k_1\ldots k_9, i_1i_2i_3} ,\ T_{j_1\ldots j_9,k_1\ldots k_9,  i_1\ldots i_6} ,\
T_{j_1\ldots j_9, k_1\ldots k_9, i_1\ldots i_8,j } ,\ \ldots 
\eqno(5.2.7)$$
where 
$$
T_{ij} = K^k{}_j\eta_{ki} + K^k{}_i\eta_{kj}, T_{i_1i_2i_3}= R^{j_1j_2j_3}\eta_{i_1j_1} \eta_{i_2 j_2} \eta_{i_3j_3} + R_{i_1i_2i_3}, \ \ldots 
\eqno(5.2.8)$$
We recall that the index range in these equations is $i,j,\ldots =2,\ldots 10$. A natural choice is to take the irreducible representation of $I_c(E_9)$ to be the one given in equation (5.2.7) except that now we regard the elements of the representation to be our fields evaluated for our chosen brane charges $l^{(0)}_A$. Hence our chosen irreducible representation contains the fields
$$
\psi(0)=\{ h_{ij}(0)= h_{(ji)}(0), \ A_{i_1i_2i_3}(0) ,\ A_{i_1\ldots i_6} (0),\  A_{i_1\ldots i_8,j }(0) ,\  \ A_{j_1\ldots j_8, i_1i_2i_3} (0), 
$$
$$
A_{j_1\ldots j_9, i_1\ldots i_6} (0),\ \ldots \} ,\ \  i, j , \ldots = 2,\ldots , 10
\eqno(5.2.9)$$
\par
 Recalling our light cone analysis of section three we recognise the field for gravity $h_{ij}$ and the three form $A_{i_1i_2i_3} $ and their duals of all  kinds. It is straightforward to verify  that it is $I_c(E_9)$ invariant to set $h_{i}{}^i=0$ and to impose the duality relations of section three at low levels. 
 Surely there are an infinite number of  $I_c(E_9)$ invariant duality relations connecting all the higher level fields to    $h_{ij}$ and the three form $A_{i_1i_2i_3} $. This is consistent with the fact that the equations of motion that follow from the $E_{11}\otimes_S l_1$ non-linear realisation, 
 found in   references [3], [4] and [18],   only contain these degrees of freedom. Putting these equations of motion in   light cone notation will lead to the duality relations we are discussing here. Thus we may take the bosonic degrees of freedom to be contained in the fields 
 $$
h_{ij} , \ A_{i_1i_2i_3} , \ \ i,j,\ldots =  2,\ldots , 10
 \eqno(5.2.10)$$
the degree of freedom count of the above  fields  is $45-1=44$ plus 84 giving 128 and they represent a graviton and a three form. Thus we  have constructed an irreducible representation of $I_c(E_{11})\otimes_s l_1$ that possess only the bosonic degrees of freedom of eleven dimensional supergravity. 
 \par
Since there are many dualities  there are many ways, that is, set of fields  to represent the 128 degrees of freedom. It is interesting to find a set that transforms as a representation of $I_c(E_{8})= SO(16)$. Such a representation is given by 
$$
h_{ij} , \ A_{i_1i_2i_3} , \  A_{i_1\ldots i_6} , \        h_{i_1\ldots i_8 , j} , \        \ \ i,j,\ldots =  3,\ldots , 10
 \eqno(5.2.11)$$
which contribute the required 128 degrees of freedom as follows $36+56+28+8=128$ . They transform in the $128=2^7$ Majorana-Weyl  spinor representation of $I_c(E_{8})=SO(16)$. We note that this choice involves the level three dual graviton field. 
\par
As explained above we require an irreducible representation  of an algebra that comes from deleting  nodes one and two  in the $E_{11}$ Dynkin diagram and then taking the  Cartan involution invariant subalgebra. This results in  the algebra $I_c(E_9)$ as shown in the Dynkin diagram below. 
$$
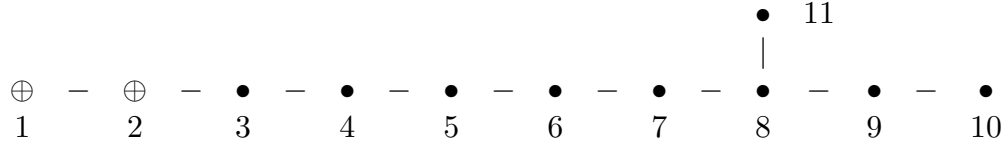
\matrix{
& & & & & & & & & & & & & & \bullet  & 11 & & & \cr 
& & & & & & & & & & & & & & | & & & & \cr
\oplus& - & \oplus & - & \bullet & - & \bullet & - & \bullet & - & \bullet & - & \bullet & - & \bullet & - & \bullet & - & \bullet \cr
1 & & 2 & & 3 & & 4 & & 5 & & 6 & & 7 & & 8 & & 9 & & 10 \cr
}
$$
where the symbol $\oplus$ indicates that the node has been deleted. 
\par
In E  the theory in $D$ dimensions is given by deleting the node labelled $D$ in the $E_{11}$ Dynkin diagram. We denote this node as $\otimes$. Thus to find the eleven dimensional theory we delete node eleven. In our present context this means we decompose $I_c(E_9)$ into $I_c(A_8)= SO(9)$
The corresponding Dynkin diagram is shown in the figure below
$$
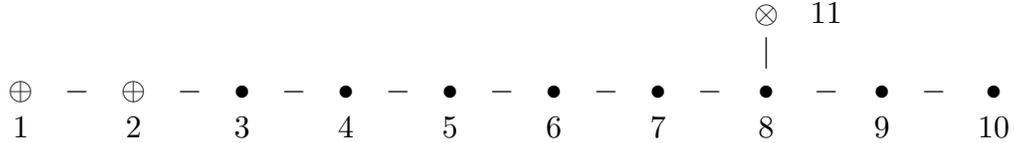
\matrix{
& & & & & & & & & & & & & & \otimes  & 11 & & & \cr 
& & & & & & & & & & & & & & | & & & & \cr
\oplus& - & \oplus & - & \bullet & - & \bullet & - & \bullet & - & \bullet & - & \bullet & - & \bullet & - & \bullet & - & \bullet \cr
1 & & 2 & & 3 & & 4 & & 5 & & 6 & & 7 & & 8 & & 9 & & 10 \cr
}
$$
In fact this is what we did above in equation (5.2.9). 
\par
To find the theory in  three dimensions we must delete node three and decompose  as indicated in the Dynkin diagram below 
$$
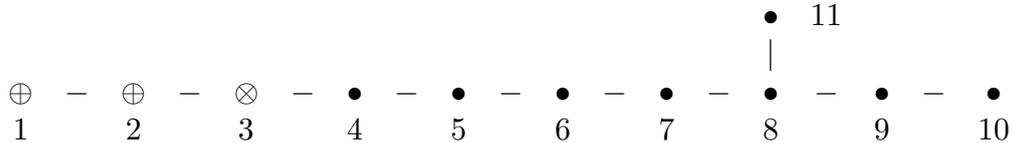
\matrix{
& & & & & & & & & & & & & & \bullet  & 11 & & & \cr 
& & & & & & & & & & & & & & | & & & & \cr
\oplus& - & \oplus & - & \otimes & - & \bullet & - & \bullet & - & \bullet & - & \bullet & - & \bullet & - & \bullet & - & \bullet \cr
1 & & 2 & & 3 & & 4 & & 5 & & 6 & & 7 & & 8 & & 9 & & 10 \cr
}
$$
In this case we must decompose $I_c(E_9)$ representation of equation (5.2.9) into  multiplets of $I_c(E_8)$.  The degrees of freedom are just those of equation (5.2.11),  which make up 128 scalars that belong to a multiplet of $I_c(E_8)=SO(16)$, as well as  the other elements in the $I_c(E_9)$ representation  of equation (5.2.9) which are related by duality relations to those of equation (5.2.11). We recall that is three dimensions gravity has no degrees of freedom. We observe that the dual graviton in eleven dimensions contributes 8 of these degrees of freedom.  Thus the degrees of freedom in the three  dimensional theory do not come in the obvious way from the lowest level eleven dimensional fields as one might naively expect. 
\par
In four  dimensions we delete node four  and so we decompose the representations of $I_c(E_9)$ into representations of $SO(2)\otimes I_c(E_7)$ as indicated in the Dynkin diagram below
$$
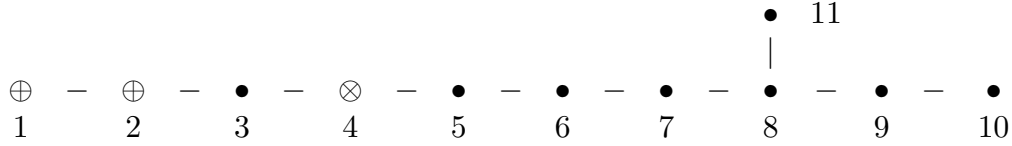
\matrix{
& & & & & & & & & & & & & & \bullet  & 11 & & & \cr 
& & & & & & & & & & & & & & | & & & & \cr
\oplus& - & \oplus & - & \bullet & - &  \otimes& - & \bullet & - & \bullet & - & \bullet & - & \bullet & - & \bullet & - & \bullet \cr
1 & & 2 & & 3 & & 4 & & 5 & & 6 & & 7 & & 8 & & 9 & & 10 \cr
}
$$
Carrying out the decomposition  of the fields in equation (5.2.9) into $SO(2)\otimes I_c(E_7)$  we find the $SO(2)\otimes I_c(E_7)$ multiplet 
$$
h_{(i^\prime j^\prime)} (2)  ;  \ h_{(ij)}  (28) ,\ A_{i_1i_2i_3} (35) ,\  A_{i_1\dots i_6} (7) ,\ ;\  h_{i^\prime j} (7) , A_{i^\prime j_1j_2} (21)
\eqno(5.2.12)$$
where now $i^\prime , j^\prime = 2,3, \ i,j,\ldots = 4, \ldots , 10$. The numbers in brackets indicate the dimension in the $I_c(E_7)$  representation. We find $70= 28+35+7$ scalars, $56=28.2= (21+7).2$ vectors and the 2 spin which all together  make up the required 128 degrees of freedom. The other fields in equation (5.29) are related to these by duality transformations. We note for example  that 
$A_{i^\prime j^\prime i}\propto   \epsilon _{i^\prime j^\prime} \epsilon_{ik_1\ldots k_6} A_{ k_1\ldots k_6}$. 
 The 128 states in the four dimensional theory are contained in equation  (5.2.12) with the other states in .equation (5.2.9) being  related to these states by duality relations  
 \par
 To find the theories in  three and four dimension we have  decomposed the degrees of freedom of equation (5.2.9)  into $I_c(E_{8})=SO(16)$ and $SO(2)\otimes I_c(E_7)$ respectively as illustrated by  the above Dynkin diagrams  and   their   indicated node deletions. 
Clearly the states in equations (5.2.11) and (5.2.12)  involve  fields that arise from different places in the eleven dimensional  theory and its irreducible representation of equation (5.2.9). Hence the 128 degrees of freedom in the three and four dimensions have different origins and so appear differently. In particular the $I_c(E_{8})$ multiplet of equation (5.2.11) involves the dual graviton while this field does not appear in the $SO(2)\otimes I_c(E_7)$ multiplet of equation (5.2.12). 
\par
We will now discuss how the three  dimensional states of equation (5.2.11) are related to the four dimensional states in equation (5.2.12) in more detail. The internal symmetry  indices of the former states can take eight values and while the latter can only take seven values and so to find the latter from the former we restrict the range in the former to the seven values and write explicitly the additional index 3 which is a spacetime index in four dimensions. The 70 scalars in four dimensions arise from the states of equation (5.2.11) as the fields 
$h_{ij}$, $A_{i_1i_2i_3} $ and $A_{i_1\ldots i_6}$ as $70=28+35 +7$ respectively where now the indices run $i,j,\ldots = 4,\ldots , 10$. 
The 56 vector degrees of freedom in four dimensions arise from the  states of equation (5.2.11) as follows. 
We find 28 states with the spacetime index 3 from the fields $h_{3i}$ and $A_{3ij}$. However, we find another 28 states also with a spacetime index 3 from the states $A_{3i_1\ldots i_5}$ and $h_{3i_1\ldots i_7, k}$. These latter states are  by the duality relations the same as lower level states, namely $A_{3i_1\ldots i_5}\propto \epsilon _{32}\epsilon_{ i_1\ldots i_5j_1j_2} A_{2 j_1j_2}$ and $h_{3i_1\ldots i_7, k}\propto \epsilon _{32}\epsilon_{i_1\ldots i_7, k}h_{2, k}$ which have a the spacetime 2 index. Thus we find a vector with both possible spacetime indices, that is, 2 and 3 and there are 56 of them. 
 Among the three  dimensional states of equation (5.2.11) we also find the states $h_{33}$ and $h_{3i_1\ldots i_7, 3}$. The latter satisfies the gravity-dual gravity relation $h_{3i_1\ldots i_7, 3}\propto \epsilon _{32}\epsilon_{ i_1\ldots i_7}h_{2, 3}$.  Hence we find the states $h_{33}$ and $h_{23}$ and thus we find the two states of the graviton in four dimensions. 
\par
 In E theory we have an $E_{11}$ symmetry in all dimensions and  the theories in the different dimensions are found by different decompositions of $E_{11}$. As a result they  are equivalent and  one can map between the different theories [19]. The above analysis traces how the 128 bosonic degrees of freedom contained in the irreducible representation of $E_{11}\otimes l_1 $ appear in the different dimension.   Of course the content of the multiplet of equation (5.2.9)  is the same no matter how one decomposes it but the appearance of the 128 degrees of freedom that appear at the  lowest level states in the corresponding internal symmetry  representations is different.  However to reconcile the results in the different dimensions one must use the duality relations and in particular the gravity-dual gravity relation.  This is to be expected as it was observed in the very first paper on $E_{11}$ that to possess the $E_8$ symmetry one had to incorporate the dual graviton and this field was not needed for the $E_7$ symmetry. When using these relations the states in four dimensions can be reorganised to belong to a multiplet of the $I_c(E_{8})$  symmetry reflecting the fact that the four dimensional theory has an $E_8$ symmetry.
\par
These observations  make contact with the interesting papers of references [20]. These authors uses a light cone formulation of the maximal supergravity theories and found some "mysterious" symmetries. In particular they found a hidden $E_{8}$ symmetry in four dimensions and traced how this arises from the three dimensional theory by carrying out certain redefinitions. As we have seen the light cone formulation  appears naturally from the analysis of the irreducible representation of  $I_c(E_{11})\otimes_s l_1$ that is contained in  the maximal supergravity theories and it was  by using this formalism that they were able to expose a part of the $E_{11}$ symmetry. The complicated field redefinitions they use should correspond to the change of variables to the dual fields required to get from the three to the four dimensional theory. As we have explained the symmetry is much more apparent if one uses the formulation that has the fields and their duals and in particular the gravity and its dual gravity field. 
It would be relatively straight forward to put the  equations of motion of the  $E_{11}\otimes l_1 $ non-linear realisation into light cone gauge and then of course one would even find more of the  $E_{11}$ symmetry depending on which level one computed to. 
 \par
 The full irreducible representation is found  by boosting using  equation (4.2), namely 
$$
\psi(\varphi)= U(e^{\varphi^{-i} J_{-i} +\varphi^{+-} J_{+-} + \varphi^{-ij} S_{-ij}+\ldots })\psi(0)
\eqno(5.2.13)$$
and the fields defined on  the generalised spacetime are given by 
$$
\int D\varphi e^{x^A D_A{}^-(e^{\varphi^{-i} J_{-i} +\varphi^{+-} J_{+-} + \varphi^{-ij} S_{-ij}+\ldots })\sqrt 2 m}\psi(\varphi)
\eqno(5.2.14)$$ 
This will automatically obey equation (4.9)  with $L^2(0)=0$ and the conditions of equations (4.11). 
We note that $\psi(\varphi)$ has a very non-trivial and completely specified dependence on the coordinates beyond those of our usual spacetime. 
\par
The final  step is to find an $I_c(E_{11})$ covariant formulation of the above irreducible  representation by embedding it in a representation of $I_c(E_{11})$. The obvious representation to embed it in is the representation of  $I_c(E_{11})$ that  provided by  the Cartan involution odd generators of $E_{11}$. Writing the elements of this representation as fields we have  
$$
A_{\underline \alpha} =\{ h_{ab}= h_{(ab)}, \ A_{a_1a_2a_3} ,\ A_{a_1\ldots a_6} ,\  A_{a_1\ldots a_8,b } ,\  \ A_{b_1\ldots b_9, a_1a_2a_3}, \ \ 
$$
$$
A_{b_1\ldots b_{10}, ( a_1a_2)}, A_{b_1\ldots b_{11},  a} ,\ A_{b_1\ldots b_9, a_1\ldots a_6} ,\ldots \}
\eqno(5.2.15)$$
where now $a,b,\ldots =0,1,\ldots ,10$. The higher level fields  can be read off from the generators in the $E_{11}$ algebra. 
\par
To embed the representation of equation (5.2.7) into that of equation (5.2.15) we have to carry out the analogous steps that we did for the massless spin one particle of the Poincare algebra  of equation (4.14). When the fields are evaluated for our chosen brane charges  we should set all the fields of equation (5.2.15) with a $+$  and a $-$ index to zero. When we boosted the first of these restrictions  will become an $I_c(E_{11}$ covariant condition which is given by 
$$
K^{AB} G_{A, B}{}^C=K^{AB} (D^{\underline \alpha})_B{}^C \partial_A A_{\underline \alpha}=0 
\eqno(5.2.16)$$
where $G_{A, B}{}^C= E_A{}^\Pi E_B{}^\Lambda \partial_\Pi E_{\Lambda} {}^C$, where $K^{AB}$ is the $I_c(E_{11}$ invariant metric  and $E_{\Pi} {}^C$ is the vielbein on the generalised spacetime which is constructed from the non-linear realisation. In the right-hand side of the equation we have linearised the condition as required by our discussion. This condition was put forward as a gauge choice and its components at low levels were worked out in detail in reference [21]. It is the analogue of $\partial^a A_a=0$ for the Poincare algebra case. 
\par
The condition of equation (5.2.16) effectively reduces the index range of the fields of equation (5.2.12) by one. To recover the irreducible representation of equation (5.2.9) we must also further reduce the index range by another one by demanding an equivalence relation. After the boost and the transformation to functions of the generalised spacetime this equivalence relation can be taken as 
$$
A_{\underline \alpha}(x^A)\sim   A_{\underline \alpha}(x^A)+  (D_{\underline \alpha}+ D_{-\underline \alpha} )_A {}^B \partial_B \Lambda^A (x^A)
\eqno(5.2.17)$$
This gauge transformation was proposed in reference [30] and evaluated for the  low level fields. This corresponds to the equivalence relation, or gauge symmetry, in the Poincare algebra case $A_a \sim A_a+ \partial_a \Lambda $
\par
We note that under the equivalence relation the condition of equation (5.2.17) changes as 
$$
K^{AB} G_{A, B}{}^C\sim K^{AB} G_{A, B}{}^C + K^{AB} (D^{\underline \alpha})_B{}^C \partial_A (D_{\underline \alpha}+ D_{-\underline \alpha} )_E {}^F \partial_F \Lambda^E 
$$
$$
\propto K^{AB} \partial_A \partial _B \Lambda^C= L^2  \Lambda^C=0
\eqno(5.2.18)$$
where in the second line we used  using equation (4.9).  This means that the constraint of equation (5.1.16) does not depend on the equivalence class, as it must do. This is the analogue in the Poincare algebra case of the condition $\partial ^a \partial_a \Lambda=0$ in the gauge parameter $\Lambda$. 
\par
The 128 degrees of freedom in the irreducible representation must be the ones that are contained in the equations of motion that follow from the $E_{11}\otimes _s l_1$ non-linear realisation. These equations of motion  were constructed in references [3] , [4] and [18] up to level four and the degrees of freedom found up to this level were just those of eleven dimensional supergravity. The fact that the irreducible representation considered in this section has the degrees of freedom of eleven dimensional supergravity and no more strongly suggests that the full equations of motion of the non-linear realisation will contain just these degrees of freedom. 
\par
In the above we embedded  the irreducible representation  into a  covariant formulation. This requires projection conditions and an equivalence relation just like the spin one irreducible representation of the Poincare algebra. However, these conditions are different to   the equations of motion both for the spin one particle (Maxwell theory) or in the equations of motion of the $E_{11}\otimes _s l_1$ non-linear realisation. It would be interesting to see if the irreducible representation provides some hints on how to systematically construct the later equations of motion. 
\medskip
{\bf 5.3 The M2 brane }
\medskip
The work in this section was carried out in collaboration with Keith Glennon. 
We now consider the charges that occur for the M2 brane, that is, $P_{\underline a} = (m,0,\ldots,0)$ and $Z^{12} = em$,  all other charges being zero. The Casimir of equations (4.8) vanishes if $e^2=1$ and so do the conditions of equation (4.11). As such  it can be thought of as  "half BPS". In this section we take the indices to have the ranges $ \underline a , \underline b,\ldots = 0,1,\ldots , 10$, $a,b, \ldots = 0,1,2$ and $a^\prime , b^\prime ,\ldots = 3, \ldots , 10$. We can write the later choice as $Z^{ij} = \epsilon^{ij} em$ where $i,j, \ldots =1,2$ here and below. The $I_c(E_{11}) $ transformations of the brane charges  are given in  appendix A and  for the above  values of the charges they become  
$$
\delta P_{\underline b} = 2m \Lambda^0{}_{\underline b}  + 6 em \Lambda_{{\underline b} 12}, \ 
\delta Z^{ \underline b_1 \underline b_2} = - 6 m \Lambda^{0 \underline b_1 \underline b_2} - 4em \Lambda^{[\underline b_1 |}{}_i \epsilon^{ij} \delta _j ^{| \underline b_2]},
\eqno(5.3.1)$$
$$
\delta Z^{\underline b_1 \ldots \underline b_5} = 360 m \Lambda^{0 \underline b_1 \ldots \underline b_5} - 60em \epsilon^{ij} \delta_{ij}^{[\underline b_1 \underline b_2} \Lambda^{\underline b_3 \underline b_4 \underline b_5]},\ 
\eqno(5.3.2)$$
$$
\delta Z^{\underline b_1 \ldots \underline b_8} = {9 \cdot 6720 \over 8} m \Lambda^{\underline b_1 \ldots \underline b_8,0} + 2520 em \epsilon^{ij} \delta _{ij} ^{[\underline b_1 \underline b_2} \Lambda^{\underline b_3 \ldots \underline b_8]}
\eqno(5.3.3)$$
$$
\delta Z^{\underline b_1 \ldots \underline b_7,\underline d} = 5670 m ( e\Lambda^{[\underline  b_1 \ldots \underline b_5 |\underline d|} \epsilon ^{ij} \delta_{ij}^{\underline b_6 \underline b_7]} - e \Lambda^{[\underline b_1 \ldots \underline b_6} \epsilon^{ij} \delta_{ij}^{\underline b_7]\underline d} -  m \Lambda^{\underline b_1 \ldots \underline b_7 \underline d, 0} )- 8.7560 m \Lambda^{0 \underline b_1 \ldots \underline b_7,\underline d}.
\eqno(5.3.4)$$
The transformations that preserve the chosen values of the charges are found by setting these variations to zero. This results in the conditions 
$$
\Lambda^{0 i} = 0, \ \Lambda^{a b^\prime} = -{3 \over 2e} \epsilon^{a c_1 c_2} e \Lambda_{c_1c_2} b{}^{b'}, \ \Lambda^{ij}\not=0 , 
, \ \Lambda^{a^\prime_1 a^\prime_2}\not=0 , 
\eqno(5.3.5)$$ 
$$
\Lambda^{a_1 a_2 a_3} = 0,  \Lambda^{a^\prime_1 a^\prime_2 0}=0 , \ \Lambda^{a^\prime_1 a^\prime_2 i}\not=0 , \ 
 \Lambda^{b_1' b_2' b_3'}= -{10\over e} \epsilon_{a_1 a_2 a_3} \Lambda^{a_1 a_2 a_3b_1' b_2' b_3'},\  
\eqno(5.3.6)$$  
$$
\Lambda^{a_1' \ldots a_6'} = - {28 \over e} \varepsilon_{b_1 b_2 b_3} \Lambda^{b_1 b_2 b_3 a_1' \ldots a_5',a_6'} = {14 \over e} \varepsilon_{b_1 b_2 b_3} \Lambda^{b_1 b_2 a_1' \ldots a_6',b_3}, 
$$
$$
\   \Lambda^{0 a_1' \ldots a_5'} = 0, \ \Lambda^{i a_1' \ldots a_5'} = {56\over 6e} \epsilon_{b_1b_2b_3}\Lambda^{b_1b_2b_3i a_1' \ldots a_5', i} , \ 
\eqno(5.3.7)$$
$$
 \ \Lambda^{0 i a_1' \ldots a_4'} = 0, \  \Lambda^{ij a_1' \ldots a_4'} \not=0 ,\ 
\eqno(5.3.8)$$
Consistency of these  equations involving $e$ requires  $e^2 = 1$ and we chose  $e=1$ 
\par
From equations (5.3.5-8) determine the non-zero parameters and substituting this into the $I_c(E_{11})$ group element of equation (1.6) 
we find the subalgebra that  preserves our above choice of charges. It is given by  
$$
{\cal H}= \{J_{ij}, J_{a^\prime_1 a_2^\prime} , \  \hat{L}_{a_1 a_2'} , \ S_{a_1^\prime a_2^\prime i}, \ 
\hat{S}_{b_1' b_2' b_3'} ,\  
\hat {S}_{i a^\prime_1 a^\prime _2 \ldots a^\prime_5 }  , \ \hat {S}_{a^\prime_1 a^\prime_2 \ldots  a^\prime_5 a^\prime_6} , \ 
  \hat{S}_{a^\prime_1\ldots a^\prime_6}, \ \ldots \}
\eqno(5.3.9)$$
where 
$$
 \hat{L}_{a_1 a_2'} = 2 J_{a_1 a_2'} +  \epsilon_{a_1}{}^{e_1 e_2} S_{e_1 e_2 a_2'}, \
 \hat{S}_{b_1' b_2' b_3'} = S_{b_1' b_2' b_3'} + {1 \over 3} \varepsilon^{e_1 e_2 e_3} S_{e_1 e_2 e_3 b_1' b_2' b_3'} , \ 
$$
$$
 \hat{S}_{i a^\prime _1 .. a^\prime_5} =   S_{i a^\prime _1 .. a^\prime_5} -{1\over 2}   \varepsilon^{e_1 e_2 e_3}  S_{i a^\prime _1 .. a^\prime_5 e_1e_2, e_3} , \ 
 $$
 $$
\hat{S}_{a^\prime _1 \ldots a^\prime _6} = S_{a^\prime_1\ldots  a^\prime_6} -{1\over 2}   \varepsilon^{e_1 e_2 e_3}  S_{ a^\prime _1 .. a^\prime_6 e_1e_2, e_3} 
\eqno(5.3.10)$$
The generators of $I_c(E_{11}) $ not in ${\cal H}$ can be chosen to be 
$$
J_{0i} ,\ J_{a^\prime b^\prime} ,\ S_{a_1a_2a_3} ,\ S_{i a^\prime_1\ldots a^\prime_5} ,\ S_{i ja^\prime_1\ldots a^\prime_4} ,\ldots  
\eqno(5.3.11)$$
\par
The algebra of ${\cal H}$ is given by 
$$
[L_{ab^\prime}, L_{cd^\prime} ]=0 ,\ [L_{ab^\prime}, \hat S_{c^\prime_1 c^\prime_2 c^\prime_3  }]=0 ,\ 
$$
$$
[L_{ab^\prime}, \hat S_{ic^\prime_1 c^\prime_2 } ]=-2 \eta_{ai} \hat S_{b^\prime c^\prime_1 c^\prime_2} + 2\epsilon_{aic} \eta_{b^\prime  [c^\prime_1} L^e{}_{c^\prime_2 ]}
 ,\ [ \hat S_{ a^\prime_1 a^\prime_2 a^\prime_3  },  \hat  S_{b^\prime_1 b^\prime_2 b^\prime_3  }]=2\hat S_{ a^\prime_1 a^\prime_2 a^\prime_3  b^\prime_1 b^\prime_2 b^\prime_3  } , 
 $$
 $$
 [S^{ia^\prime_1a\prime_2}, S_{j b^\prime_1 b^\prime_2} ]= 2S^{ia^\prime_1a\prime_2}{}_{j b^\prime_1b\prime_2}
 -2\delta ^{a^\prime_1a\prime_2} _{b^\prime_1b\prime_2}    J^i{}_j       -4 \delta^i_j \delta ^{[ a^\prime_1}_{[ b^\prime_1} J^{a^\prime_2}{}_{b^\prime_2]} ,\ 
 \ldots 
 \eqno(5.3.12)$$
 together with the usual commutators for $J_{0i}$ and $J_{a^\prime b^\prime} $. The second commutator in this last equation corrects an error in reference [13]. 
 \par
 The next step is to choose an irreducible representation of ${\cal H}$ and implement section four. We leave this to a future work. 

\medskip
{\bf 5.4 The M5 brane  }
\medskip
The work in this section was carried out in collaboration with Keith Glennon. 
We now consider the charges that occur for the M5 brane, that is, $P_{\underline a} = (m,0,\ldots,0)$ and $Z^{12345} = em$,  all other charges being zero. The Casimir of equations (4.8) vanishes if $e^2=1$. The conditions of equation (4.11) also vanish   and so it can be thought of as  ``half BPS". In this section we take the indices to have the ranges $ \underline a , \underline b,\ldots = 0,1,\ldots , 10$, $a,b, \ldots = 0,1, \ldots , 5$ and $a^\prime , b^\prime ,\ldots = 6, \ldots , 10$. We can write the latter choice as $Z^{i_1 \ldots i_5} = \epsilon^{i_1 \ldots i_5} em$ where $i_1, i_2, \ldots =1,\ldots , 5$ and we also set $p = 1, \ldots, 4$ here and below. For the values of the charges given above, the $I_c(E_{11}) $ transformations of the brane charges given in appendix A  become 
$$
\delta P_{\underline{b}} = 2 m ( \Lambda^0{}_{\underline{b}} - 180 e \Lambda_{\underline{b} 1 \ldots 5}), \ \  \delta Z^{\underline{b}_1 \underline{b}_2} = \delta^{\underline{e}_1 \underline{e}_2 \underline{e}_3 \underline{b}_1 \underline{b}_2}_{i_1 \ldots i_5} \varepsilon^{i_1 \ldots i_5} em \Lambda_{\underline{e}_1 \underline{e}_2 \underline{e}_3}  - 6m \Lambda^{0 \underline{b}_1 \underline{b}_2}, \eqno(5.4.1)$$
$$
\delta Z^{\underline{b}_1 \ldots \underline{b}_5} = 10 m (36 \Lambda^{0 \underline{b}_1 \ldots \underline{b}_5} - e \varepsilon^{i_1 \ldots i_5} \Lambda^{[\underline{b}_1}{}_{\underline{c}} \delta^{|\underline{c}| \underline{b}_2 \ldots \underline{b}_5]}_{i_1 \ldots i_5}) , \eqno(5.4.2)$$
$$
\delta Z^{\underline{b}_1 \ldots \underline{b}_8} = 42 m (180  \Lambda^{\underline{b}_1 \ldots \underline{b}_8,0} + e \varepsilon^{i_1 \ldots i_5} \Lambda^{[\underline{b}_1 \underline{b}_2 \underline{b}_3} \delta^{\underline{b}_4 \ldots \underline{b}_8]}_{i_1 \ldots i_5}) , \eqno(5.4.3)$$
$$
\delta Z^{\underline{b}_1 \ldots \underline{b}_7,\underline{d}} = - {945 \over 4} [  em \varepsilon^{i_1 \ldots i_5} ( \delta^{[\underline{b}_1 \ldots \underline{b}_5}_{i_1 \ldots i_5} \Lambda^{\underline{b}_6 \underline{b}_7]\underline{d}} +  \delta^{[\underline{b}_1 \ldots \underline{b}_4 |\underline{d}|}_{i_1 \ldots i_5} \Lambda^{\underline{b}_5 \underline{b}_6 \underline{b}_7]})  + 32 ( \Lambda^{\underline{b}_1 \ldots \underline{b}_7\underline{d},0}  + 8 m \Lambda^{0 \underline{b}_1 \ldots \underline{b}_7,\underline{d}})]. 
\eqno(5.4.4)$$
The transformations that preserve the chosen values of the charges are found by setting these variations to zero. This results in the conditions
$$
\Lambda^{0i} = 0, \ \Lambda^{i_1 i_2} \neq 0, \  \Lambda^{a b'} = - {3 \over 2 e} \varepsilon^{a}{}_{c_1 \ldots c_5} \Lambda^{c_1 \ldots c_5 b'}, \ \Lambda^{a' b'} \neq 0, \eqno(5.4.5) $$
$$
\Lambda^{a_1 a_2 a_3} = {e \over 3!} \varepsilon^{a_1 a_2 a_3 b_1 b_2 b_3} \Lambda_{b_1 b_2 b_3}, \ \Lambda^{0 i a'} = 0, \ \Lambda^{i_1 i_2 a'} \neq 0, \ \Lambda^{0 a_1'a_2'} = 0, \ \eqno(5.4.6)$$
  $$
 \Lambda^{i a_1' a_2'} = -{14 \over 3 e} \varepsilon_{b_1 \ldots b_6} \Lambda^{b_1 \ldots b_6 a_1' a_2', i}, \ \Lambda^{a^\prime_1 a^\prime_2 a^\prime_3} = - {7\over e} \varepsilon_{b_1 \ldots b_6} \Lambda^{b_1 \ldots b_6 a^\prime_1 a^\prime_2 , a^\prime_3}  \eqno(5.4.8) $$
 Consistency of these conditions involving $e$ requires  and we choose $e = 1$. Inserting these conditions into the $I_c(E_{11})$ group element of equation (1.6) we find the subalgebra that preserves our above choice of charges. It is given by 
$$
{\cal H} = \{ J_{i_1 i_2}, \ J_{a_1' a_2'}, \  L_{a_1 a_2'}, \ S_{-a_1 a_2 a_3}, \ S_{i_1 i_2 a_3'}, \ \hat{S}_{i a_2' a_3'}, 
\ \hat{S}_{a_1' a_2' a_3'}, \ S_{i_1 \ldots i_4 a_5' a_6'}, 
$$
$$
\ S_{i_1 i_2 i_3 a_1' a_2' a_3'}, \  S_{i_1 i_2 a_1' \ldots a_4'}, 
\ S_{i a_1' \ldots a_5'},  \ S_{a_1' \ldots a_6'},
\eqno(5.4.9)$$
where 
$$
L_{a_1 a_2'} = 2 ( J_{a_1 a_2'} + {2 \over 5!} e \varepsilon_{a_1}{}^{c_1 \ldots c_5} S_{c_1 \ldots c_5 b'}), \ 
S_{\pm a_1 a_2 a_3} = {1 \over 2} (S_{a_1 a_2 a_3} \pm {1 \over 3!} e \varepsilon_{a_1 a_2 a_3 b_1 b_2 b_3} S^{b_1 b_2 b_3}), $$
$$
\hat{S}_{i a_2' a_3'} = S_{i a_2' a_3'} -{1\over 5!} e \varepsilon^{b_1 \ldots b_6} S_{ b_1 \ldots b_6 [a^\prime_2  a^\prime_3 , i]}, $$
$$
\hat{S}_{a_1' a_2' a_3'} = S_{a_1' a_2' a_3'} -{1\over 5!} \varepsilon^{b_1 \ldots b_6}  S_{b_1 \ldots b_6 [a_1' a_2',a_3']} 
. \eqno(5.4.10)$$ 
The generators of $I_c(E_{11})$ not in ${\cal H}$ are thus
$$
J_{0i}, \ S_{a_1 a_2 a_3}, \ S_{0 a b'}, \ S_{0 a_1' a_2'}, \ S_{a_1 \ldots a_6}, \ S_{0 a_1 a_2 a_3 b_1' b_2'}, S_{0 a_1 a_2 b_1' b_2' b_3'}, \ S_{0 a b_1' \ldots b_4'}, \ S_{0 a_1' \ldots a_5'} , 
\eqno(5.4.11)$$
\par
We leave it to a future work to find an irreducible representation of this algebra these generators obey  following the discussion of  section four. 

\medskip
{\bf 6 A discussion on the generalised spacetime}
\medskip
It is very widely believed that spacetime must be replaced  in a fundamental theory of strings and branes by some deeper concept. The low energy effective action of strings and branes possess an $E_{11}$ and it is the non-linear realisation of  $E_{11}\otimes_s l_1$. The $E_{11}$ symmetry implies that in addition to the usual coordinates of spacetime this theory has an infinite number of spacetime coordinates. In fact the $E_{11}$ invariance of the equations of motion depends  crucially on the higher level coordinates and as a result these coordinates are not some decoration but are essential for the way the theory works. This is  a strong sign that they play a role which has some deeper physical meaning. It has been suggested  [3] that the generalised spacetime  in the non-linear realisation is a kind of effective spacetime. We are very familiar with the idea of a  low energy  effective field theory which  determines the behaviour of the underlying  theory at low energies even though it may have very different fields, and even concepts,  to those in the underlying theory from which it can be derived. Similarly, we should think of the infinite dimensional spacetime as  an effective spacetime  in E theory that encode effects of an underlying theory in which spacetime has been replaced with something else. 
\par
The maximal supergravity theories were formulated around forty years ago and  we have got used to the fact that most of them  they live in spacetimes that are larger than the spacetime with four dimensions in which we obviously  live. We have also grown to accept the spacetimes of twenty six and ten dimensions in which  the bosonic string and the superstrings  live respectively. Although we have by now got very used to these higher dimensional  spacetimes, there is a very substantial reluctance to consider spacetimes that are significantly different to these.  Indeed  if a theory has more coordinates than these spacetimes there has been an understandable desire to find mechanisms that eliminate the additional coordinates.  However, it is good to remember that there is no experimental evidence that we live in ten, or eleven dimensions and furthermore, there are no very reliable predictions from supersymmetry, or string theory,  that could be tested even if we had the ability to carry out experiments at much higher energies that we have at our disposal at present. Hence when working in the context of E theory, it is better to simply accept the $E_{11}$ prediction that we  live in an infinite dimensional spacetime rather than try to find ways to  get rid of the additional coordinates. 
\par
An analogous situation occurred with brane charges in E theory which are infinite in number and  belong to the vector representation of  $E_{11}$. 
While the supersymmetry algebra in eleven dimensions has only, the translations,  a two form and  five form central charges,  the vector representation  begins with the translations and  then contains in order of increasing level the two and five forms and  then an infinite number of further charges. However, it is not correct to  dismiss the higher order charges found in the vector representation as it is clear that they have a physical interpretation that is needed in the underlying theory. To give one example,    the next brane charge up is associated with  the Taub-Nut solution. The   elements in the vector representation contain all the known brane charges and one can expect it contains  all possible   brane charges.   In fact the coordinates arise in the non-linear realisation from the generators in the vector representation that correspond to the brane charges and so there is a close connection between the brane charges and the infinite coordinates in the spacetime in E theory. This strongly suggests that if one keeps the brane charges one should also keep all the coordinates. 
\par
The role of spacetime is to uniquely label the  events that occur and up until relatively recently this was assumed to be the collision of point particles. However, we now believe that we have strings and  branes,  indeed $E_{11}$ predicts an infinite number of such objects. As is often stated strings do not part,  or join,  at given points in spacetime as this point is observer dependent. A  point particle has just its mass and spin, which are associated with the irreducible representation of the Poincare algebra to which it belongs, as well as certain quantum numbers  associated with internal symmetry groups. However, extended objects can have more structure. One  example is provided by strings which can have  addition winding modes.  Yet other examples are the world volume fields living on for example D branes and five branes. 
This   suggests that our usual approach to spacetime does not hold when we have  extend objects. In this section we will now argue that  the  spacetime encoded in E theory does appear to  take account of the additional structures that  occur for  extended objects. One encouraging sign is that in the usual approach to brane dynamics the embedding coordinates of the brane and the fields describing the world volume fields are very different origins. However, in the approach to brane dynamics  in E theory there is a unified approach in that the embedding coordinates and  the world volume  fields both appear as coordinates that belong to the  vector representation, [12,13,14]. 
\par
In what follows we will point out some similarities between the strings and branes that occur  in E theory and a much studied extended object, namely the monopole,  which is a solution of the $N=4$ supersymmetric Yang-Mills theory as well as many other  gauge theories that are spontaneously broken. We now very briefly recall some of the main features of monopoles.  The magnetic charge and mass of the monopole are  $Q_m=-{4\pi\over e}$ and $M_m=({4v\pi\over e})^2$ respectively where  $e$ is the coupling in the Yang-Mills theory and $v$ the expectation value of the scalar field in the theory.  This solution has four moduli, its three positions $x^i$ in space and one additional coordinate $\chi$ which arises from a large gauge transformation, that is, one that is non-trivial at infinity. 
\par
At low energy the  motion of the monopole is described by letting the moduli depend on time, the Manton approximation [23]. The   action which describes this motion  is found by taking the moduli in the  classical solution to depend on time and then substituting this into the action of the field theory in which the monopole solution was found, say  the $N=4$ supersymmetric Yang-Mills action. The motion in the coordinate $\chi$ is related to fact that the monopole can also carry an electric charge, that is, be a dyon. If we quantise the action we find the usual momenta ${\bf p}=-i{\partial \over \partial {\bf x}}$and the momenta corresponding to the coordinate $\chi$ is given by  $Q=-i{\partial \over \partial \chi}$. A momentum eigenstate has the form 
$e^{i{\bf x}\cdot {\bf p}}e^{in\chi}$ where $n$ is an integer as the coordinate $\chi$ takes the range $0\le \chi\le 2\pi$. The electric charge being $Q_e= ne$.  
\par
For monopoles that occur in extended supersymmetric theories we have the additional structure of    the supersymmetry algebra which for  $N=2$ theories takes the form 
$$
\{ Q^i_\alpha . Q^j_\beta \}= 2 (\gamma^a C)_{\alpha \beta} P_a \delta^{ij} +\epsilon ^{ij} v^2((Q_e +\gamma_5 Q_m)C)_{\alpha \beta}
\eqno(6.1)$$
We note that we can write the supersymmetry as a semi-direct product, namely 
${\rm Lorentz}\otimes_s {\cal T}$ where ${\cal T}$ contains the translations and the supercharges. 
\par
The two central charges in equation (6.1) must be  conserved charges and can be identified with the electric and magnetic charges of the monopole.  If we demand that we have short super multiplets, then the supersymmetry algebra implies that the monopoles satisfy   BPS conditions of the form 
$$
p^2+ v^2(Q_m^2 +Q_e^2) =0
\eqno(6.2)$$
A detailed account of monopoles can be found in the review of reference [24] and the  reviews in the book [25].
\par
\par
Long ago [26] and [27] it was argued that the many features of the monopoles that occur in the $N=4$ supersymmetric Yang-Mills theory can be seen as arising from the uplift of this theory to six dimensions. There the central charges that occur in the  supersymmetry algebra of equation (6.1) in four dimensions appear as the components of the momenta in the fifth and sixth directions [26, 27]. It was shown  using the supersymmetry algebra that the correct electric and magnetic charges appeared in this way giving them a six dimensional origin [26] and the topological nature of these charges was further analysed in reference [27] using the compact nature of the additional directions. These  results show  that there is a  considerable simplification when the theory is lifted  to six dimensions. 
\par
If we were to second quantise the  action describing the monopole motion resulting from the Manton approximation, discussed above, we would find a relativistic quantum field theory whose fields $\psi$ would depend on the usual coordinates of spacetime but also on the coordinate $\chi$. One might expect that it would obey, at the linearised level, a condition of the very generic  form  
$$
(-\partial_a \partial^a + v^2 e^2{\partial^2\over \partial\chi^2}  + ({4\pi v\over e})^2 )\psi=0
\eqno(6.3)$$
This equation should describe not only the monopoles but also the dyons for a suitable set of fields $\psi$. One could also add another coordinate $\rho$ corresponding to the monopole charge to find a quantum field theory in six dimensions four of which are the usual coordinates and Minkowski spacetime and two additional coordinates $\chi$ and $\rho$. When the dependence on these  extra coordinates is absent 
we would  expect the usual $N=4$ supersymmetric theory describing the elementary particles. However, we might expect the theory to contain the monopoles and also the dyons when the additional coordinates $\chi$ and $\rho$ are active. In this case the duality transformations would act in the theory and in particular on the additional coordinates which belong to a torus. In addition, since a quantum field theory contains operators that create many particles one may also hope that this theory  will also contain the scattering of several monopoles and so the corresponding moduli space. It would be of interest to explore these ideas further. 
\par
One can view the appearance of the above additional coordinates beyond those of our usual spacetime  from another perspective. We note that  the supersymmetry algebra is the semi-direct product 
${\rm Lorentz}\otimes_s {\cal T}$ where ${\cal T}$ contains the translations and the supercharges. 
The  superspace corresponding to the algebra of equation (6.1) is the coset of the supersymmetry algebra with the  local subgroup  being the Lorentz algebra. 
The group element will have the generic form 
$$
g= e^{x^a P_a + z vQ_e +w vQ_m} e^{\theta^{\alpha i} Q_{\alpha i}}
\eqno(6.4)$$
The superfields will depend on the coordinates $x^a, z, w, \theta^{\alpha i}$. If we work in a formulation in which the $\theta^{\alpha i}  $ coordinates have been eliminated by integrating  over them in the action then we will be left with  fields that depend on $x^a, z, w$. Carry out a transformation $g_0=e^{aQ_e}$ on the group element of equation (6.4)   would shift the $z$ coordinate $z \to z+a$. As a result we can interpret the coordinate $z$  as the moduli coordinate $\chi$. This argument also suggest that one should also include   the other coordinate $w$ associated with the magnetic charge and identified with $\rho$. Hence from the superspace view point the uplift to a six dimensional space is very natural. We note that the monopole solution was found in the $N=4$ Yang-Mills theory that did not have these additional coordinates but they are required when one wants a second quantised theory that describes monopoles. 
\par
One can also see the dependence on the additional coordinates from the more  basic  perspective of the irreducible representations of the supersymmetry algebra of equation (6.1). Following the discussion of the irreducible representations given in this paper,  applied to the supersymmetry algebra,  we can find  fields in the extended  momentum space which have  values for the usual momenta $p_a$ as well as the additional  "momenta" $Q_e$ and $Q_m$. Making the transition to 
 the fields that depend on the corresponding enlarged spacetime one would find that the factor in the analogue of equation (4.6) would contain the term 
 $$
 e^{x^a P_a(0) + z Q_e(0)+wQ_m(0) + \ldots }= e^{{4\pi v\over e} x^0 + evn z+{4\pi v \over e} w \ldots }
 \eqno(6.5)$$ 
  Thus we find that  fields in spacetime  do depend on the extra coordinates and that this dependence is a straight forward  consequence of the construction of the irreducible representations.  This is an inevitable consequence of the fact that the supersymmetry algebra combines the electric and magnetic charges in the same algebra as the usual translations and the Lorentz group. 
  \par
We will now apply some of the lessons learnt in the context of the monopole to the theory of strings and branes. The scheme set out above for the monopole  was as follows:  a general  brane solution of the field theory has  moduli  which arise from  large  gauge transformations in the field theory, These moduli become  the coordinates  which describe  the    motion of the brane. The resulting second quantised  quantum field theory has fields that depend on the coordinates of the brane motion. As a result  we find  that the internal structure of the monopole results in a field theory that lives in an extended spacetime whose additional coordinates arise from large gauge transformations that belong to the part of the gauge symmetry that is not spontaneously broken. 
\par 
  A similar picture to that which merged for the monopole  should hold  for the solutions of  the low energy effective field theory of strings and branes and in particular the maximal supergravity theories. Such     solutions, that is,  brane solutions  have moduli which are associated with large gauge transformations of the field theory. Furthermore, the low energy motion of the corresponding branes  can, at least in  principle, be  found by using a generalisation of the Manton method. From this approach one can  recover not only the embedding of the branes as they move in the background spacetime but also the the dynamics of any world volume fields the brane may possess, see reference [22] in this context. 
 \par
One can also construct brane dynamics  from the different view point, namely it can be derived from the  $E_{11}\otimes_s l_1$ non-linear realisation [12,13,14].  To do this  one constructs the non-linear realisation of $E_{11}\otimes_s l_1$  with {\bf a local subalgebra that is a  subalgebra of $I_c(E_{11})$}. The different choices of subalgebra lead to the different branes and by taking different decompositions of $E_{11}$ one finds the branes in the different dimensions.  A consequence of  the way the non-linear realisation works is that the coordinates arise as parameters in the group element associated with the vector representation and so they are in a one to one correspondence with the elements of the vector representation and as a result the coordinates that describe the brane motion are in one to one correspondence with the brane charges. These coordinates are taken to depend on the parameters that describe the world volume of the brane. These coordinates describe not only the embedding coordinates of the brane as it moves through spacetime but also the dynamics  of  any world volume fields  the brane may possess. 
\par
 If we comparing this approach with the approach of Manton applied to branes and discussed just above we would suspect that the coordinates of the vector representation can be thought of  arising from  the moduli of brane  solutions which are in turn  associated with large gauge transformations of the low energy effective field theory.  Let us now examine if these statements could be true.  
 \par
We recall that the low energy effective field theory of strings and branes is the non-linear realisation of $E_{11}\otimes_s l_1$  with {\bf  the local subalgebra being $I_c(E_{11})$}.  This theory has an infinite number of fields associated with $E_{11}$ which depend on  an infinite number of coordinates which are associated with the vector ($l_1$) representation.  The fields are in a one to one correspondence with the generators which correspond to a positive roots,  essentially the Borel subalgebra  of $E_{11}$ and they depend on the coordinates.  When restricted to the low level fields and usual coordinates of spacetime  these, essentially unique, equations are those of  the maximal supergravity theories [3,4].  The non-linear realisation mentioned above that is used to construct the branes looks very similar but the local subgroup is different and in the brane case   the coordinates  depend on the  parameters of the brane world volume.  As we have mentioned,  the way the $E_{11}\otimes_s l_1$ non-linear realisation is constructed means that the coordinates of the generalised spacetime are in one to one correspondence with the members of the vector representation. The same coordinates describe the motion of the brane  and appear in the field theory. This is indeed what we would expect if the coordinates in the field theory arise from the brane coordinates which in turn resulted from the moduli of the solution. 
\par
   We will now examine to what extend these coordinates could arise from large gauge transformations.    At low levels the fields in the non-linear realisation, that is, the low energy effective action of strings and branes, are in a one to one correspondence but at higher levels there is associated usually more than one element  of the vector representation to each field. 
However,   every gauge symmetry of the fields is in a one to one relation with the elements of the vector representation [30]. Hence  as the  coordinates are   in the vector representation they are   in a one to one correspondence with the gauge transformations and so it is reasonable to assume that the coordinates have their origin in large gauge transformations. 
\par
In the table below we illustrate the fact that, at low levels,  the brane charges, the coordinates in the field theory and  the brane dynamics and  the gauge transformations of the fields  are all in one to one correspondence.  
 $$
 \matrix {{\rm field} &h_a{}^b & A_{a_1a_2a_3} & A_{a_1\ldots a_6}&\ldots  \cr
 {\rm  gauge \ transformation  }&\xi^a & \Lambda_{a_1a_2} & \Lambda_{a_1\ldots a_5}&\ldots\cr 
\  {\rm generator\   in \ vector \ representation }& P_a & Z^{a_1a_2}     &Z^{a_1\ldots a_5}&\ldots\cr  
{\rm coordinate\  in \ field \ theory\  or\  brane\  dynamics} & x^a & x_{a_1a_2}     &x_{a_1\ldots a_5}&\ldots\cr  }
  \eqno(6.6)$$
\par
We observe that the above connection between  the "Manton" approach and the non-linear realisation approaches  to brane dynamics  holds in the sense that the parameters in the large gauge transformations do match up with  the coordinates in the brane dynamics.  This encourages the belief that  the  coordinates in the generalised spacetime that occurs in the field theory could  be seen as  arising from  large gauge transformations associated with the brane charges. 
 \par
We will now compare the way  the additional coordinates that arise in the context of the irreducible representations that describe the monopole and those that arise in the study of branes. The former is an irreducible representation of the supersymmetry algebra and the dependence of the extra coordinates $z$ and $w$ are given in equation (6.5). The branes arise as irreducible representations of $I_c(E_{11})\otimes_s l_1$ and we have explained how to construct them in section four.  In particular in equation (4.6) we formulated  the dependence of the fields of the irreducible representation on the generalised spacetime and in section five we gave some examples of this dependence for specific irreducible representations. We see that the dependence of the fields on the higher level coordinates is  non-trivial and determined. Examining equation (4.6) we see that the dependence on the generalised spacetime in encoded in the factor 
$$
e^{x^A D_A{}^B (e^{\varphi\cdot S}) l_B (0)}= e^{x^a P_a(0) + x_{a_1a_2} Z^{a_1a_2}(0)+\ldots }
\eqno(6.7)$$
Comparing with equation (6.5) we conclude that the additional coordinates arise in the same generic way. 
\par
One can view this from a more general view point. The algebra $E_{11}$, like supersymmetry, combines spacetime symmetries such as the Lorentz algebra with  internal symmetries. Hence unlike for the Poincare algebra in which the irreducible representations (particles) are labelled only by their mass and spin,   the irreducible representations of $I_c(E_{11})\otimes_s l_1$ (branes) contain states that carry mass and spin but also quantum numbers of the  internal symmetries which are contained in $E_{11}$.  Hence $E_{11}$ unifies the internal symmetry charges with the  mass and spin. In fact $E_{11}$ is another way, apart from supersymmetry,  of overcoming the well known no go theorem of Coleman and Mandula. 
A result of this unification is that  the space time fields the occur in the irreducible representations of  $I_c(E_{11})\otimes_s l_1$ depend on 
 additional coordinates associated with the brane charges. This is consistent with the general view expressed above that additional coordinates are required when we are describing  objects which are extended and have internal structure. 
\par
We can summarise these considerations as follows 
\medskip
{\bf  The $E_{11}$ symmetry  unifies the  Poincare   algebra  with other   symmetries which in dimensions lower than eleven  include  internal symmetries.  As a result the irreducible representations of $I_c(E_{11})\otimes_s l_1$  describe  branes which  carry quantum numbers associated with the  Poincare algebra and these additional symmetries. It follows that branes  move in a generalised spacetime that has coordinates  in addition  to those of our  usual spacetime. }
\medskip
There is another aspect that monopoles and branes have in common. Crucial to the discovery of monopoles were the BPS conditions which were proposed in the context of theories which were not supersymmetric. In E theory  branes also obey the analogue of BPS conditions, for example those of equations (4.9) and (4.11). Indeed  the on shell conditions in x spacetime of equation (4.10) are very similar to the condition of equation (6.2) for the monopole in that they lead to conditions between the usual momenta and the internal  charges. In both contexts the BPS conditions, when subject to considerable restrictions in the E theory case, can  be seen as  the BPS conditions that arise from the supersymmetry algebra by demanding that one has short representations. Yet another similarity is that  both theories possess duality symmetries, indeed large parts of $E_{11}$ are duality symmetries. We hope to expand on these subjects elsewhere. 
\par
The rigid transformations of $E_{11}$ lead to transformations of the coordinates and fields. For example the infinitesimal transformation $g_0=e^{a_{b_1b_2b_3}R^{b_1b_2b_3}}$ leads at lowest order to the transformations 
$$
\delta x_{a_1a_2}= 3 a_{a_1a_2c}x^c\equiv 3\Lambda_{a_1a_2} , \ \ \delta A_{a_1a_2a_3}= a_{a_1a_2a_3}=\partial_{[a_1}\Lambda_{a_2a_3]}
\eqno(6.8)$$
among others. We observe that if we write the transformation of  $ x_{a_1a_2}$  as the parameter $ \Lambda_{a_1a_2}$ then the transformation of $A_{a_1a_2a_3}$ can be written as a gauge transformation with this gauge parameter, albeit one that is  of a very restricted form. In fact this is a quite general result that applies to all the fields and rigid transformations of the $E_{11}\otimes_s l_1$ non-linear realisation. Thus the usual gauge transformation at least in restricted from  is part of the non-linear realisation.  The $E_{11}\otimes_s l_1$ non-linear realisation does not explicitly have as symmetries the full gauge and  diffeomorphisms symmetries  that we are familiar with. However, it turns out that the, essentially unique, equations of motion, when computed to the relevant degree in derivatives do have these  symmetries. The reason for this is not clear at present. 
If by slight of hand we were to identify $\Lambda_{a_1a_2}$ with the corresponding coordinate $x_{a_1a_2}$ then we could write the last of the transformations  in equation (6.8) as $\delta A_{a_1a_2a_3}= a_{a_1a_2a_3}=\partial_{[a_1}x_{a_2a_3]}$ which although a very crude argument is consistent with the above proposition that the coordinates in the generalised spacetime arise due to the large gauge transformations. We note that seen in  this way these considerations  includes the usual coordinates of spacetime. 
\par
We also note that the shift in the field in equation (6.8) corresponds to the fact that the fields in the non-linear realisation are Goldstone fields corresponding to the spontaneous braking of the $E_{11}$ symmetry and so part of the gauge symmetry can be viewed as arising due to spontaneous symmetry breaking of a rigid symmetry, perhaps the part related to large gauge transformations. The same applies to the coordinates of the generalised spacetime, at least as seen from the perspective of brane dynamics. 
\par
Of course the theories we have been discussing do contain  the usual coordinates of spacetime as part of their definition. However, the underlying theory of strings and branes will have spacetime replaced by some other concept but  one can imagine that it will have some analogue of gauge and diffeomorphism symmetries. Such "local" symmetries that are not spontaneously  broken  will appear as the local symmetries and we  speculate  that  their  large gauge transformations  give rise to the generalised spacetime  in the  $E_{11}\otimes_s l_1$ non-linear realisation including the  spacetime we are used to. 

\medskip
{\bf {Conclusions }}
\medskip
In this paper we have explained how to construct the Poincare algebra analogue of particles in E theory, that is,  we have, at least  in principle,  constructed the irreducible representations of $I_c(E_{11})\otimes_s l_1$ which describe branes.   The starting point for this construction is to  choose a  set of  brane charges that are  non-zero and find the subalgebra that preserves this choice. This choice may  satisfy the analogue of BPS conditions which can lead to shorter multiplets.  Even for  a given choice of non-zero brane charges  there are a very many irreducible representations. We have constructed one  irreducible representation   in detail and shown that it contains only the degrees of freedom of eleven dimensional supergravity. The radical reduction in this irreducible representation 
is due to the existence of an infinite number of invariant duality relations. In order to analyse these we used  a light cone formulation of duality relations. 
\par
The equations of motion that results from the non-linear realisation of $E_{11}\otimes_s l_1$ with local subgroup $I_c(E_{11})$ have been constructed up to level four and have been found to contain the degrees of freedom of eleven dimensional supergravity as on-shell states [3,4,18].  It must be that these equations of motion encode the above mentioned  irreducible representation  and this reinforces the belief that these are the only degrees of freedom that these equations contain. 
It is inevitable  that other irreducible representations will generically contain an infinite number of degrees of freedom and it would be interesting to understand these in more detail. This avenue of investigation considerable increases the possible applications of E theory. 
\par
The light cone formalism analysis of the above  irreducible representation makes contact with the work of reference [20] that uses the light cone formalism to find a "mysterious" $E_8$ symmetry in four dimensional maximal supergravity. We have argued  that this symmetry is part of the $E_{11}$ that has been exposed by the use of the light cone formalism. These authors have also discussed the lift of the Eulers symmetry of three dimensional gravity. This will be a part of the $A_1^{+++}$ symmetry of gravity in four  dimensions [28, 21] in an almost identical way and the same arguments we given in this paper will apply when suitably restricted to this context. It would be of interest to explore this in greater detail. One way to proceed would be to put the $E_{11}$ equations of motion in the light cone formulation. 
\par
We have also argued that the additional coordinates found in the generalised spacetime that arises in the non-linear realisation of $E_{11}\otimes_s l_1$ are a consequence of the fact that $E_{11}$ unifies the usual spacetime symmetries with  "internal symmetries". Indeed we have argued that extended objects with internal structure always require such additional coordinates and this includes the well known monopole. 
We have also argued that  spacetime arise from the analogue of  large gauge transformations in some underlying theory in which the concept of spacetime is replaced by something else.  
\par
As we have mentioned one can construct brane dynamics from an $E_{11}\otimes_s l_1$ non-linear realisation by  using  suitable local subalgebras. The different choices of local subalgebra lead to the different branes. However, precisely what the local subalgebra one should take  for certain well known branes was deduce in reference [12,13,14] by demanding  consistency of  the resulting dynamics. However, looking at the results of this paper we see that the local subalgebra one should use in the non-linear realisation to obtain the dynamics of a given branes is essentially just the one that preserves the charges of that brane. The only difference is that one has to introduce the world volume symmetry of the brane into the subalgebra that preserves the brane charges. For example for the M2 brane rather than demand it preserve the choice of brane charges of section 5.3  we can instead demand that it preserve the condition $Z^{ab}= \epsilon^{abc}P_c$. This observation should allow the systematic construction of brane dynamics using the non-linear realisation,  at least in principle.

\medskip
{\bf {Acknowledgements}}
\medskip
We wish to thank Keith Glennon for his collaboration on the work reported in sections  5.3, 5.4 and appendix A, Nicolas Boulanger for discussions on gravity duals which is to appear in our forthcoming paper,  Nic Manton for discussions on monopoles and Dionysios Anninos for general discussions.  We also wish to thank the SFTC for support from Consolidated grants numbers ST/J002798/1 and ST/P000258/1.

\medskip
{\bf Appendix A. The $I_c(E_{11})$ variation of the brane charges}
\medskip
The work in this section was carried out in collaboration with Keith Glennon. In this paper we require the variations of the brane charges under the Cartan involution invariant subalgebra of $E_{11}$ which are given by
$$
\delta l_A = [\Lambda^{\underline{\alpha}} S_{\underline{\alpha}},l_A(0)] = 0
\eqno(A.1)$$

Using the algebra of the $E_{11}\otimes_s l_1$  found in [31],  and earlier $E_{11}$ publications. the $l_1$ variations are found to be
$$
\delta P_{\underline{b}} = 2 \Lambda^{\underline{a}_1}{}_{\underline{b}} P_{\underline{a}_1} + 3 \Lambda_{\underline{b} \, \underline{c}_2 \underline{c}_3}   Z^{\underline{c}_2 \underline{c}_3} - 3  \Lambda_{\underline{b} \, \underline{c}_2 \ldots \underline{c}_6} Z^{\underline{c}_2 \ldots \underline{c}_6} - {4 \over 3} \Lambda_{\underline{c}_1 \ldots \underline{c}_8,\underline{b}} Z^{\underline{c}_1 \ldots \underline{c}_8}
$$
$$
+ {4 \over 3} \Lambda_{\underline{b} \, \underline{c}_2 \ldots \underline{c}_8,\underline{d}}  Z^{\underline{c}_2 \ldots \underline{c}_8 \underline{d}} + {4 \over 3} \Lambda_{\underline{b} \, \underline{c}_2 \ldots \underline{c}_8,\underline{d}}  Z^{\underline{c}_2 \ldots \underline{c}_8 , \underline{d}} + \ldots
\eqno(A.2)$$
$$
\delta Z^{\underline{b}_1 \underline{b}_2} = - 4 \Lambda^{[\underline{b}_1}{}_{\underline{c}}Z^{|\underline{c}|\underline{b}_2]} + \Lambda_{\underline{c}_1 \underline{c}_2 \underline{c}_3} Z^{\underline{c}_1 \underline{c}_2 \underline{c}_3 \underline{b}_1 \underline{b}_2} -  6 \Lambda^{\underline{b}_1 \underline{b}_2 \underline{a}_3} P_{\underline{a}_3} - \Lambda_{\underline{c}_1 \ldots \underline{c}_6} Z^{\underline{c}_1 \ldots \underline{c}_6 \underline{b}_1 \underline{b}_2}
$$
$$
+ {1 \over 3} \Lambda_{\underline{c}_1 \ldots \underline{c}_6} Z^{\underline{c}_1 \ldots \underline{c}_6 [\underline{b}_1, \underline{b}_2]} - {16 \over 135} \Lambda_{\underline{c}_1 \ldots \underline{c}_8,\underline{d}} Z^{\underline{b}_1 \underline{b}_2 \underline{d} \, \underline{c}_1 \ldots \underline{c}_5,\underline{c}_6 \underline{c}_7 \underline{c}_8} + {4 \over 63} \Lambda_{\underline{c}_1 \ldots \underline{c}_8,\underline{d}} \hat{Z}^{\underline{b}_1 \underline{b}_2 \underline{c}_1 \ldots \underline{c}_7,\underline{c}_8 \, \underline{d}} $$
$$
+ {16 \over 189} \Lambda_{\underline{c}_1 \ldots \underline{c}_8,\underline{d}} Z^{\underline{b}_1 \underline{b}_2 \underline{d} \, \underline{c}_1 \ldots \underline{c}_6,\underline{c}_7 \underline{c}_8}  - {16 \over 189} \Lambda_{\underline{c}_1 \ldots \underline{c}_8,\underline{d}} Z^{\underline{b}_1 \underline{b}_2 \underline{c}_1 \ldots \underline{c}_7,\underline{c}_8\underline{d}} + {1 \over 42} \Lambda_{\underline{c}_1 \ldots \underline{c}_8,\underline{d}} Z_{(1)}^{\underline{b}_1 \underline{b}_2 \underline{d} \, \underline{c}_1 \ldots \underline{c}_7,\underline{c}_8} $$
$$
- {1 \over 42} \Lambda_{\underline{c}_1 \ldots \underline{c}_8,\underline{d}} Z_{(1)}^{\underline{b}_1 \underline{b}_2 \underline{c}_1 \ldots \underline{c}_8,\underline{d}} - {1 \over 6} \Lambda_{\underline{c}_1 \ldots \underline{c}_8,\underline{d}} Z_{(2)}^{\underline{b}_1 \underline{b}_2 \underline{d} \, \underline{c}_1 \ldots \underline{c}_7,\underline{c}_8} + {1 \over 6} \Lambda_{\underline{c}_1 \ldots \underline{c}_8,\underline{d}} Z_{(2)}^{\underline{b}_1 \underline{b}_2 \underline{c}_1 \ldots \underline{c}_8 , \underline{d}} + \ldots $$
$$
\delta Z^{\underline{b}_1 \ldots \underline{b}_5} = - 10 \Lambda^{[\underline{b}_1}{}_{\underline{c}} Z^{|\underline{c}| \underline{b}_2 \ldots \underline{b}_5]} + \Lambda_{\underline{c}_1 \underline{c}_2 \underline{c}_3} Z^{\underline{b}_1 \ldots \underline{b}_5 \underline{c}_1 \underline{c}_2 \underline{c}_3 } + \Lambda_{\underline{c}_1 \underline{c}_2 \underline{c}_3} Z^{\underline{b}_1 \ldots \underline{b}_5 \underline{c}_1 \underline{c}_2 , \underline{c}_3} - 60 Z^{[\underline{b}_1 \underline{b}_2} \Lambda^{\underline{b}_3 \underline{b}_4 \underline{b}_5]}$$
$$
- \Lambda_{\underline{c}_1 \ldots \underline{c}_6} Z^{\underline{c}_1 \ldots \underline{c}_6 \underline{b}_1 \ldots \underline{b}_5} + {4 \over 189} \Lambda_{\underline{c}_1 \ldots \underline{c}_6} Z^{\underline{c}_1 \ldots \underline{c}_6 [\underline{b}_1 \underline{b}_2 \underline{b}_3 , \underline{b}_4 \underline{b}_5]}  - {40 \over 441} \Lambda_{\underline{c}_1 \ldots \underline{c}_6} Z^{\underline{c}_1 \ldots \underline{c}_6 [\underline{b}_1 \underline{b}_2 \underline{b}_3,\underline{b}_4 \underline{b}_5]} $$
$$
- {55 \over 336} \Lambda_{\underline{c}_1 \ldots \underline{c}_6} Z_{(1)}^{\underline{c}_1 \ldots \underline{c}_6 [\underline{b}_1 \ldots \underline{b}_4,\underline{b}_5]}  + {5 \over 16} \Lambda_{\underline{c}_1 \ldots \underline{c}_6} Z_{(2)}^{\underline{c}_1 \ldots \underline{c}_6 [\underline{b}_1 \ldots \underline{b}_4,\underline{b}_5]} - 360 \Lambda^{\underline{b}_1 \ldots \underline{b}_5 \underline{a}_6} P_{\underline{a}_6} + \ldots 
\eqno(A.3)$$
$$
\delta Z^{\underline{b}_1 \ldots \underline{b}_8} = - 16 \Lambda^{[\underline{b}_1}{}_{\underline{c}} Z^{|\underline{c}| \underline{b}_2 \ldots \underline{b}_8]} + \Lambda_{\underline{c}_1 \underline{c}_2 \underline{c}_3} Z^{\underline{c}_1 \underline{c}_2 \underline{c}_3 \underline{b}_1 \ldots \underline{b}_8} + {4 \over 135} \Lambda_{\underline{c}_1 \underline{c}_2 \underline{c}_3} Z^{\underline{c}_1 \underline{c}_2 \underline{c}_3 [\underline{b}_1 \ldots \underline{b}_5,\underline{b}_6 \underline{b}_7 \underline{b}_8]} 
$$
$$
- {20 \over 63} \Lambda_{\underline{c}_1 \underline{c}_2 \underline{c}_3} Z^{\underline{c}_1 \underline{c}_2 \underline{c}_3 [\underline{b}_1 \ldots \underline{b}_6,\underline{b}_7 \underline{b}_8]} - \Lambda_{\underline{c}_1 \underline{c}_2 \underline{c}_3} Z_{(2)}^{\underline{c}_1 \underline{c}_2 \underline{c}_3 [\underline{b}_1 \ldots \underline{b}_7,\underline{b}_8]} + 42 \Lambda^{[\underline{b}_1 \underline{b}_2 \underline{b}_3} Z^{\underline{b}_4 \ldots \underline{b}_8]}$$
$$
+ 2520 \Lambda^{[\underline{a}_1 \ldots \underline{a}_6} Z^{\underline{b}_7 \underline{b}_8]} + {9 \cdot 6720 \over 8} \Lambda^{\underline{b}_1 \ldots \underline{b}_8,\underline{e}} P_{\underline{e}} + \ldots
\eqno(A.4)$$
$$
\delta Z^{\underline{b}_1 \ldots \underline{b}_7,\underline{d}} = - 14 \Lambda^{[\underline{b}_1}{}_{\underline{c}} Z^{|\underline{c}| \underline{b}_2 \ldots \underline{b}_7],\underline{d}} - 2 \Lambda^{\underline{d}}{}_{\underline{c}} Z^{\underline{b}_1 \ldots \underline{b}_7,\underline{c}} + \Lambda_{\underline{c}_1 \underline{c}_2 \underline{c}_3} Z^{\underline{c}_1 \underline{c}_2 \underline{c}_3 \underline{d} [\underline{b}_1 \ldots \underline{b}_4,\underline{b}_5 \underline{b}_6 \underline{b}_7]} 
$$
$$
+ {1 \over 2} \Lambda_{\underline{c}_1 \underline{c}_2 \underline{c}_3} Z^{\underline{d}[\underline{b}_1 \ldots \underline{b}_5 |\underline{c}_1 \underline{c}_2,\underline{c}_3| \underline{b}_6 \underline{b}_7]} $$
$$
+ \Lambda_{\underline{c}_1 \underline{c}_2 \underline{c}_3} Z^{\underline{c}_1 \underline{c}_2 \underline{c}_3 \underline{d} [\underline{b}_1 \ldots \underline{b}_5,\underline{b}_6 \underline{b}_7]} - {3 \over 7} \Lambda_{\underline{c}_1 \underline{c}_2 \underline{c}_3} Z^{\underline{d}[\underline{b}_1 \ldots \underline{b}_6 | \underline{c}_1 \underline{c}_2 , \underline{c}_3 | \underline{b}_7]} - \Lambda_{\underline{c}_1 \underline{c}_2 \underline{c}_3} \hat{Z}^{\underline{d}[\underline{b}_1 \ldots \underline{b}_6 |\underline{c}_1 \underline{c}_2,\underline{c}_3|\underline{b}_7]} $$
$$
-  \Lambda_{\underline{c}_1 \underline{c}_2 \underline{c}_3} Z_{(1)}^{\underline{c}_1 \underline{c}_2 \underline{c}_3 \underline{d} [\underline{b}_1 \ldots \underline{b}_6,\underline{b}_7]} - {3 \over 8}  \Lambda_{\underline{c}_1 \underline{c}_2 \underline{c}_3} Z_{(1)}^{\underline{d} \underline{b}_1 \ldots \underline{b}_7 \underline{c}_1 \underline{c}_2,\underline{c}_3} - {945 \over 4} Z^{[\underline{b}_1 \ldots \underline{b}_5}\Lambda^{\underline{b}_6 \underline{b}_7] \underline{d}} $$
$$
- {945 \over 4}  Z^{[\underline{b}_1 \ldots \underline{b}_4|\underline{d}|}\Lambda^{\underline{b}_5 \underline{b}_6 \underline{b}_7]} + 5670 \Lambda^{[\underline{b}_1 \ldots \underline{b}_5 |\underline{d}|} Z^{\underline{b}_6 \underline{b}_7]} - 5670 \Lambda^{[\underline{b}_1 \ldots \underline{b}_6} Z^{\underline{b}_7]\underline{d}}  $$
$$
- 7560 \Lambda^{\underline{b}_1 \ldots \underline{b}_7 \underline{d} ,\underline{a}_1}  P_{\underline{a}_1} - 60480 \Lambda^{\underline{a}_1 \underline{b}_1 \ldots \underline{b}_7,\underline{d}} P_{\underline{a}_1} + \ldots
\eqno(A.5)$$
where $+\ldots $ means higher order terms in the parameters and the charges.

\medskip
{\bf {Appendix B Irreducible representations of $SO(D,D)\otimes T^{2D}$}}
\medskip
We will now outline the construction of the irreducible representations of the algebra $SO(D,D)\otimes T^{2D}$ whose algebra is given by 
$$
[K^{\underline a}{}_{\underline b},K^{\underline c}{}_{\underline d}]=\delta _{\underline b}^{\underline c} K^{\underline a}{}_{\underline d} - \delta _d^{\underline a} K^{\underline c}{}_{\underline b}, \ \
[K^{\underline a}{}_{\underline b}, R^{{\underline c} {\underline d}}]=\delta_{\underline b}^{\underline c} R^{{\underline a}{\underline d}}-\delta_{\underline b}^{\underline d} R^{{\underline a}{\underline c}},
[K^{\underline a}{}_{\underline b}, \tilde R_{{\underline c} {\underline d}}]=-\delta_{\underline c}^{\underline a} \tilde R_{{\underline b}{\underline d}}+\delta_{\underline d}^{\underline a} \tilde
R_{{\underline b}{\underline c}},
$$
$$
[R^{a{\underline b}}, \tilde R_{{\underline c} {\underline d}}]=\delta_{[{\underline c}}^{[a}
K^{{\underline b}]}{}_{{\underline d}]},\ \ \ [R^{a{\underline b}}, R^{{\underline c} {\underline d}}]=0=[\tilde R_{a{\underline b}}, \tilde R_{{\underline c} {\underline d}}]
\eqno(B.1)$$
$$
[K^{\underline c}{}_{\underline b}, P_{\underline a}]=-\delta_{\underline a}^{\underline c} P_{\underline b},\ \ [ R^{{\underline a}{\underline b}}, P_{\underline c}]=-{1\over 2}
(\delta^{\underline a}_{\underline c} Q^{\underline b}-\delta^{\underline b}_{\underline c} Q^{\underline a}),\ \ [\tilde R_{{\underline a}{\underline b}}, P_{\underline c}]=0,
\eqno(B.2)$$
$$
[K^{\underline a}{}_{\underline b}, Q^{\underline c}]=\delta_{\underline b}^{\underline c} Q^{\underline a},\ \ [ \tilde R_{{\underline a}{\underline b}}, Q^{\underline c}]={1\over
2} (\delta_{\underline a}^{\underline c} P_{\underline b}-\delta_{\underline b}^{\underline c} P_{\underline a}),\ \ [R^{{\underline a}{\underline b}},Q^{\underline c}]=0,
\eqno(B.3)$$
where $\underline a , \underline b , \ldots = 0,1,\ldots ,D-1$.
\par
The Cartan involution acts on the generators in the following way
$$ 
I_c(K^{\underline a}{}_{\underline b}) = -K^{\underline b}{}_{\underline a}, \ \ I_c(R^{{\underline a}{\underline b}}) = - \tilde{R}_{{\underline a}{\underline b}} 
\eqno(B.4)$$
and as a result the  Cartan Involution invariant subalgebra, $I_c(SO(10,10))$ is generated by

$$ 
J_{\underline{a}}{}_{\underline{b}} \equiv K^{\underline{c}}{}_{\underline{b}}\eta_{\underline{c}\underline{a}} - K^{\underline{c}}{}_{\underline{a}}\eta_{\underline{c} \underline{b}}, \ \ S_{\underline{a}\underline{b}} \equiv 2(R^{\underline{c}\underline{c}}\eta_{\underline{c} \underline{a}} \eta_{\underline{d}\underline{ b}} 
 - \tilde{R}_{\underline{a}\underline{b}})\eqno(B.5) $$
The $I_c(SO(10,10))$  algebra is given by 
$$
[J_{\underline{a} \underline{b}}, J_{\underline{c} \underline{d}}]=  
\eta_{\underline{b} \underline{c}}J_{\underline{a} \underline{d}}
-\eta_{\underline{a} \underline{c}}J_{\underline{b} \underline{d}}
-\eta_{\underline{b} \underline{d}}J_{\underline{a} \underline{c}}
+\eta_{\underline{b} \underline{c}}J_{\underline{a} \underline{d}}
$$
$$
[S_{\underline{a} \underline{b}}, S_{\underline{c} \underline{d}}]=  
\eta_{\underline{b} \underline{c}}J_{\underline{a} \underline{d}}
-\eta_{\underline{a} \underline{c}}J_{\underline{b} \underline{d}}
-\eta_{\underline{b} \underline{d}}J_{\underline{a} \underline{c}}
+\eta_{\underline{b} \underline{c}}J_{\underline{a} \underline{d}}
$$
$$
[J_{\underline{a} \underline{b}}, S_{\underline{c} \underline{d}}]=  
\eta_{\underline{b} \underline{c}}S_{\underline{a} \underline{d}}
-\eta_{\underline{a} \underline{c}}S_{\underline{b} \underline{d}}
-\eta_{\underline{b} \underline{d}}S_{\underline{a} \underline{c}}
+\eta_{\underline{b} \underline{c}}S_{\underline{a} \underline{d}}
\eqno(B.6)$$
By adopting suitable generators one sees that it is none other than the algebra $SO(10)\otimes SO(10))$. Their 
commutators with the $l_1$ representation are  given by 
    $$ [J_{\underline{a}\underline{b}}, P_{\underline{c}}] = -2\eta_{\underline{c}[\underline{a}}P_{\underline{b}]}, \ \ [S_{\underline{a}\underline{b}}, P_{\underline{c}}] = -2\delta_{\underline{c}}^{[\underline{a}}Q^{\underline{b}]} $$
$$ [J_{\underline{a}\underline{b}}, Q^{\underline{c}}] = -2\delta^{\underline{c}}_{[\underline{a}}Q_{\underline{b}]} , \ \ [S_{\underline{a}\underline{b}}, Q^{\underline{c}}] = - 2\delta^{\underline{c}}_{[\underline{a}}P_{\underline{b}]}  
\eqno(B.7)$$
\par
The brane charges transform as 
$$
\delta P_{\underline{a}}= -2\Lambda _{\underline{a}}{}^{\underline{b}}P_{\underline{b}}-2\tilde \Lambda _{\underline{a}}{}_{\underline{b}} Q^{\underline{b}} ,\ \ \delta Q^{\underline{a}}= -2\Lambda ^{\underline{a}}{}_{\underline{b}} Q^{\underline{b}}-2\tilde \Lambda ^{\underline{a}}{}^{\underline{b}} P_{\underline{b}} 
\eqno(B.8)$$
\par
Let us consider that the only non-zero values of the  brane charges are given by $P_0=m$ and $Q_1=m$ for which the $I_c(SO(10,10))$ invariant 
$\Delta \equiv P_{\underline{a}}^2+ Q_{\underline{a}}^2=0$. Examining equation (B.7) we find that these  values are preserved if 
$$\Lambda_{01}=0= \tilde \Lambda_{01} ,\ \ \Lambda_{i0}+\tilde \Lambda _{i1}=0 ,\ \ \Lambda_{i1}+\tilde \Lambda _{i0}=0 , \ \ i=2,\ldots , D-1
\eqno(B.9)$$
Hence the subalgebra that preserves our choice of brane charges is given by 
$$
{\cal H}=\{ J_{ij}, \ \ L_{i0}\equiv J_{i0}+ S_{i1}, \ \  L_{i1}\equiv  S_{i0}+J_{i1}, \ \ S_{ij} \}= \{J_{ij}, \ \ L_{i+}, \ \  L_{i-}, \ \ S_{ij} \}
\eqno(B.10)$$
where in the last equation we have rewritten the generators  using light-cone notation. 
We find that 
$$
[L_{+i},L_{+j}]=0 ,\ [L_{+i},L_{-j}]=0 ,\ [L_{-i},L_{-j}]=0 
\eqno(B.11)$$
\par
We must now choose  an irreducible  representation of ${\cal H}$. However, in view of the vanishing commutators of equation (B.10) we can take 
$L_{+i}$ and $ L_{-i}$ to vanish on the representation and as a result we are left with only the generators of $I_c(SO(D-2,D-2))$, namely  $\{ J_{ij}, S_{ij} \}$. We choose to take the irreducible  representation of $ I_c(SO(D-2,D-2))= SO(D-2)\otimes SO(D-2)$ formed from the Cartan involution odd generators of  SO(D-2,D-2), that is $T_{ij}= K_{ij}+K_{ji}, \ \tilde T_{ij}\equiv R^{kl}\delta_{ki} \delta_{lj}- \tilde R_{ij}$. These correspond to taking the fields $h_{ij}=h_{ji}, \phi, B_{ij}$. We recognise the states of the closed bosonic string, namely the graviton,  scalar and two form. 

\medskip
{\bf {References}}
\medskip
\item{[1]} P. West, {\it $E_{11}$ and M Theory}, Class. Quant.  
Grav.  {\bf 18}, (2001) 4443, hep-th/ 0104081. 
\item{[2]} P. West, {\it $E_{11}$, SL(32) and Central Charges},
Phys. Lett. {\bf B 575} (2003) 333-342,  hep-th/0307098. 
\item {[3]} A. Tumanov and P. West, {\it E11 must be a symmetry of strings and branes},  arXiv:1512.01644. 
\item{[4]} A. Tumanov and P. West, {\it E11 in 11D}, Phys.Lett. B758 (2016) 278, arXiv:1601.03974. 
\item{[5]}  P. West, A brief review of E theory, Proceedings of Abdus Salam's 90th  Birthday meeting, 25-28 January 2016, NTU, Singapore, Editors L. Brink, M. Duff and K. Phua, World Scientific Publishing and IJMPA, {\bf Vol 31}, No 26 (2016) 1630043,  arXiv:1609.06863.  
\item{[6]}  E. P. Wigner, ÒOn unitary representations of the inhomogeneous Lorentz group,Ó Annals Math.40(1939) 149.
\item{[7]}  P. West, {\it Introduction to Strings and Branes}, Cambridge University Press, 2012. 
\item{[8]} P. West, {\it Dual gravity and E11},  arXiv:1411.0920.
\item{[9]} P. West,  {\it On the different formulations of the E11 equations of motion}, Mod.Phys.Lett. A32 (2017) no.18, 1750096,  arXiv:1704.00580.
\item{[10]} N. Boulanger and P. West, forth coming publication. 
\item{[11]} N. Boulanger,  Per Sundell and  P. West,  {\it Gauge fields and infinite chains of dualities},  JHEP 1509 (2015) 192,  arXiv:1502.07909. 
\item{[12]} P. West, {\it Brane dynamics, central charges and $E_{11}$}, JHEP 0503 (2005) 077, hep-th/0412336.  
\item{[13]} P. West,  {\it E11, Brane Dynamics and Duality Symmetries}, Int.J.Mod.Phys. A33 (2018) no.13, 1850080, arXiv:1801.00669. 
\item{[14]} P. West, {\it A sketch of brane dynamics in seven and eight dimension using E theory}, Int.J.Mod.Phys. A33 (2018) no.32, 1850187, 
arXiv:1807.04176.
\item{[15]} P. West, {\it Generalised BPS conditions}, Mod.Phys.Lett. A27 (2012) 1250202, arXiv:1208.3397.
\item{[16]} N. Obers,Ê B. Pioline and E.Ê Rabinovici, {\it M-theory and U-duality on $T^d$ with gauge backgrounds}, {\tt hep-th/9712084}; N. Obers and B. Pioline,~ {\it U-duality and ÊM-theory, an algebraic approach}~, {\tt hep-th/9812139}; N. Obers and B. Pioline, {\it U-duality and ÊM-theory}, {\tt arXiv:hep-th/9809039}. 
\item{[17]} X.  Bekaert and N.    Boulanger, {\it The Unitary representations of the Poincare group in any spacetime dimension},  hep-th/061126. 
 review 
\item{[18]}  A. Tumanov and and P. West, {\it $E_{11}$,  Romans theory and higher level duality relations}, IJMPA, {\bf Vol 32}, No 26 (2017) 1750023,  arXiv:1611.03369.  
\item{[19]} P. West, {\it The IIA, IIB and eleven dimensional theories and their common $E_{11}$ origin}, Nucl. Phys. B693 (2004) 76-102, hep-th/0402140. 
\item{[20]}  S.  Ananth, L.  Brink and  S. Majumdar, {\it E8 in N=8 supergravity in four dimensions}, JHEP 1801 (2018) 024,  arXiv:1711.09110;  	S.  Ananth and L.  Brink, {\it A hidden symmetry of quantum gravity}, 10.1007/JHEP11(2018)078, arXiv:1808.0249. 
\item{[21]} M. Pettit and P. West, {\it An E11 invariant gauge fixing}, Int.J.Mod.Phys. A33 (2018) no.01, 1850009, Int.J.Mod.Phys. A33 (2018) no.01, 1850009.  
\item{[22]} T. Adawi, M.  Cederwall, U. Gran, Bengt E.W. Nilsson and B.  Razazneja. {\it  	
Goldstone tensor modes}, JHEP 9902 (1999) 001, hep-th/981114. 
\item{[23]} N. Manton, {\it A Remark on the Scattering of BPS Monopoles}, Phys.Lett. 110B (1982) 54-56, 
\item{[24]} J. Figueroa-O'Farrill, review can be found at
\par
 https://www.maths.ed.ac.uk/~jmf/Teaching/Lectures/EDC.pd. 
\item{[25]} D. Olive and P. West, {\it  Dualities and Supersymmetric Theories}, Cambridge University Press, 1999. 
\item{[26]} D. Olive,The Electric and Magnetic Charges as Extra Components of Four Momentum, Nucl.Phys. B153 (1979) 1-12
 \item{[27]} P. Zizzi, {\it An Extension of the {Kaluza-Klein} Picture for the $N=4$ Supersymmetric Yang-Mills Theory}, Phys.Lett. 137B (1984) 57-61;
 	{\it The Five-dimensional Origin Of The Monopole N=2 Supermultiplet}, Phys.Lett. 134B (1984) 197; {\i  	
A Kaluza-klein Picture Of Electric Magnetic Duality In Supersymmetry}, Nucl.Phys. B228 (1983) 229-241. 
 \item{[28]} N. Lambert and P. West, {\it Enhanced Coset Symmetries and Higher Derivative Corrections},   Phys.Rev. D74 (2006) 065002, hep-th/0603255. 
 \item{[29]}  F. Riccioni and P. West, {\it Dual fields and $E_{11}$},    Phys.Lett.B645 (2007) 286-292,  hep-th/0612001. 
 \item{[30]} P. West,  {\it  Generalised Space-time and Gauge Transformations}, JHEP 1408 (2014) 050, arXiv:1403.6395. 
 \item{[31]} N. Gromov, M. Pettit and P. West, {\it The $E_{11}$ algebra at higher levels} to appear. 
 \item{[32]}  X. Bekaert, N. Boulanger and D. Francia, {\it Mixed-symmetry multiplets and higher-spin curvatures}, arXiv:hep-th/1501.02462; 
  X. Bekaert and N. Boulanger, {\it Mixed symmetry gauge fields in a flat background},  arXiv:hep-th/0310209v2. 
  \item{[33]} X. Bekaert and N. Boulanger, {\it Tensor gauge fields in arbitrary representations ofGL(D,R):II. Quadratic actions},  arXiv:hep-th/0606198. 

\end